\documentclass[11pt,english]{article}
\usepackage{rotating}
\usepackage{latexsym}
\usepackage{graphicx}
\usepackage{psfrag}
\usepackage{amsmath,amssymb}
\usepackage{bbm}
\makeatletter
\def\section{\@startsection {section}{1}{\z@}{-3.5ex plus -1ex minus
 -.2ex}{2.3ex plus .2ex}{\large\bf}}
\def\subsection{\@startsection{subsection}{2}{\z@}{-3.25ex plus -1ex
minus -.2ex}{1.5ex plus .2ex}{\normalsize\bf}}
\makeatother
\makeatletter

\@addtoreset{equation}{section}

\makeatother

\newcommand{\captionfonts}{\small}
\makeatletter  
\long\def\@makecaption#1#2{%
  \vskip\abovecaptionskip
  \sbox\@tempboxa{{\captionfonts #1: #2}}%
  \ifdim \wd\@tempboxa >\hsize
    {\captionfonts #1: #2\par}
  \else
    \hbox to\hsize{\hfil\box\@tempboxa\hfil}%
  \fi
  \vskip\belowcaptionskip}
\makeatother   

\def\UV{{\textsc{uv}}}
\def\IR{{\textsc{ir}}}
\def\Ls{{\textsc{l}}}
\def\Rs{{\textsc{r}}}
\def\dslash{\raisebox{1pt}{$\slash$} \hspace{-6pt} \partial}
\def\pslash{\raisebox{0pt}{$\slash$} \hspace{-6pt} p}

\def\bea{\begin{eqnarray}} \def\eea{\end{eqnarray}}
\def\be{\begin{equation}} \def\ee{\end{equation}}
\def\nn{\nonumber}
  
 \def\ov{\overline}
\def\I{\mathbbm 1}
\def\ds{\displaystyle} \def\de{\partial}

\def\mc{\mathcal}
\def\ZZ{\mathbb Z}

\newcommand{\promille}{%
  \relax\ifmmode\promillezeichen
        \else\leavevmode\(\mathsurround=0pt\promillezeichen\)\fi}
\newcommand{\promillezeichen}{%
  \kern-.05em%
  \raise.5ex\hbox{\the\scriptfont0 0}%
  \kern-.15em/\kern-.15em%
  \lower.25ex\hbox{\the\scriptfont0 00}}

\begin{document}

\thispagestyle{empty}

\begin{center}
\hfill UAB-FT-629 \\
\hfill SISSA-18/2007/EP \\

\begin{center}

\vspace{1.7cm}

{\LARGE\bf Effective Action and Holography\\[3mm]
in 5D Gauge Theories}

\end{center}

\vspace{1.4cm}

{\bf Giuliano Panico$^{a}$ and Andrea Wulzer$^{b}$}\\

\vspace{1.2cm}

${}^a\!\!$
{\em ISAS-SISSA and INFN, Via Beirut 2-4, I-34014 Trieste, Italy}

\vspace{.3cm}

${}^b\!\!$
{\em { IFAE, Universitat Aut{\`o}noma de Barcelona,
08193 Bellaterra, Barcelona}}

\end{center}

\vspace{0.8cm}

\centerline{\bf Abstract}
\vspace{2 mm}
\begin{quote}\small

We apply the holographic method to $5D$ gauge theories on the warped interval. Our treatment includes the scalars associated with the fifth gauge field component, which appear as $4D$ Goldstone bosons in the holographic effective action.

Applications are considered to two classes of models in which these scalars play an important role. In the Composite--Higgs (and/or Gauge--Higgs Unification) scenario, the scalars are interpreted as the Higgs field and we use the holographic recipe to compute its one--loop potential. In $AdS/QCD$ models, the scalars are identified with the mesons and we compute holographically the Chiral Perturbation Theory Lagrangian up to $p^4$ order.

We also discuss, using the holographic perspective, the effect of including a Chern--Simons term in the $5D$ gauge Lagrangian. We show that it makes a Wess--Zumino--Witten term to
appear in the holographic effective action. This is immediately applied to $AdS/QCD$,
where a Chern--Simons term is needed in order to mimic the Adler--Bardeen chiral anomaly.

\end{quote}

\vfill

\newpage

\section{Introduction}

Models with one extra dimension compactified on an interval have been widely studied in the last years, especially after
Randall and Sundrum (RS) \cite{RandallSundrum} pointed out the virtues of
``warped'' compactifications. Applications have been considered to a wide range of physical
problems but, in view of the starting of LHC, the many warped and flat extra--dimensional
models of New Physics at the TeV scale deserve a particular mention. An incomplete list
includes Higgsless theories \cite{Csaki:2003dt,Csaki:2003zu} and models of Gauge--Higgs Unification in flat \cite{Hatanaka:1998yp,Scrucca:2003ra,Panico:2006em} and warped \cite{Contino:2003ve,Agashe:2004rs,Agashe:2005dk,Contino:2006qr} space. The latter scenario, which is also interpreted as a calculable version of Composite--Higgs \cite{GeorgiKaplan}, appears particularly promising. The models of ``holographic QCD'' (so--called $AdS/QCD$) are another remarkable application of the warped segment \cite{QCD1,QCD2}: inspired by the $AdS/CFT$ correspondence \cite{MaldacenaGubser, Witten:1998qj}, such simple
phenomenological constructions  mimic some aspects of strong interactions to a certain level
of accuracy.

The standard approach to theories on the interval is the Kaluza--Klein (KK) decomposition. One expands the $5D$ field on a basis of mass eigenstates and rewrites the theory in terms of an infinite KK tower of $4D$ fields. At low energies, an effective theory is obtained by integrating out the heavy KK states and retaining only the zero--mode. A useful alternative \cite{Luty:2003vm,Barbieri:2003pr} (for the case of fermions see \cite{Contino:2004vy})
is the so--called ``holographic'' procedure, which consists in separating the value $\widehat\Phi(x)=\Phi(x,z_{\UV})$ of the $5D$ fields at one boundary from the ``bulk'' degrees of freedom contained in $\Phi(x,z)$ for $z\neq z_{\UV}$. By integrating out the latter, one gets an effective holographic Lagrangian for the boundary field $\widehat\Phi(x)$ which is not a mass eigenstate but a linear combination of all the KK modes. As long as $\widehat\Phi(x)$ has a non--vanishing overlap with the zero--mode, however, it can safely be used to describe the lightest state and the holographic Lagrangian is a perfectly valid effective description of the original theory. For any physical process, the holographic theory will give exactly the same results as the standard effective theory for the zero--modes.

When a dual $AdS/CFT$ interpretation is possible, holography is necessary \cite{Witten:1998qj} to compute correlators of the dual $4D$ theory, because the boundary values of $5D$ fields are sources for the $4D$ operators. The holographic procedure, however, is just a technical tool which can be applied to any $5D$ theory even if no $4D$ dual exists as it is the case, for instance, when the space is flat. Though completely equivalent to the standard one, the holographic analysis is much simpler for certain applications. An example is the calculation of ``universal'' corrections induced by KK exchange \cite{Barbieri:2003pr,Barbieri:2004qk} (see also \cite{Panico:2006em,Agashe:2004rs} for the application to more realistic models). Another interesting application is the determination of three--point functions and in particular of the $Zb\bar b$ vertex \cite{Agashe:2005dk}.

Aim of the present paper is to apply the holographic method to the case of $5D$ gauge theories, we will develop a precise recipe to construct the holographic effective action for a general bulk group and general boundary conditions. The holographic action we construct includes the scalar degrees of freedom which may arise, depending on the boundary conditions, from the fifth gauge field component. As a matter of fact, these scalars play an important role in many interesting models, and an holographic effective theory in which they are included can then have very useful applications. The holographic theory contains indeed all the $4D$ fields which are needed to describe the light states, including the scalars, of the $5D$ model. It can then provide a useful low--energy effective description. 

In the (flat or warped) models of Gauge--Higgs Unification, for instance, the scalars which
come from an extended EW gauge group in $5D$ are interpreted as the Higgs field and its
dynamics could be described by the holographic effective theory which we discuss. Moreover,
when an $AdS/CFT$ interpretation is possible, the scalars correspond to (possibly
pseudo--)Goldstone bosons of a spontaneously broken global symmetry of the dual $4D$ theory
\cite{Contino:2003ve}. This is the reason why Gauge--Higgs Unification models in warped space
also describe a Composite Higgs. The simplest version of $AdS/QCD$ (discussed in detail in
\cite{Hirn:2005nr}) is also based on this observation, the mesons arise from  the fifth
component of $SU(3)_L\times SU(3)_R$ gauge fields when the bulk group is broken at the $IR$ to
its vector subgroup. In this case, the holographic effective Lagrangian is interpreted as the
one of Chiral Perturbation Theory ($\chi${\bf{PT}}) and the holographic method can be used
to extract $AdS/QCD$ predictions of the $\chi${\bf{PT}} parameters in an efficient way.

The paper is organized as follows. In section 2 we discuss the holography of non--Abelian gauge theories with general bulk group $G$ broken to $H$ at the $IR$ brane and introduce a convenient gauge--fixing. In our gauge there are no ghosts, the scalars are described by a $4D$ Goldston boson matrix which parametrizes the $G/H$ coset and the holographic action is invariant under the full bulk gauge group $G$. This makes completely explicit the $AdS/CFT$ interpretation and provides a useful tool even when this interpretation is absent.

In sect.~3 we consider applications to the Composite--Higgs (and/or Gauge--Higgs Unification) scenario. By integrating out the bulk at tree--level we derive the  holographic effective action and use it to compute the Higgs effective potential at one loop. Compared with the standard KK one, the holographic calculation is much simpler. The holographic approach is also useful for $AdS/QCD$, as we discuss in sect.~4. For such case, we determine the $\chi${\bf{PT}} action to $O(p^4)$ and we show that our result coincides with the one derived in \cite{Hirn:2005nr} after a long KK calculation.

Section 5 is devoted to more theoretical considerations. We discuss how the presence of a Chern--Simons (CS) term in the $5D$ Lagrangian affects the $4D$ holographic action. As expected \cite{Hill:2006wu}, the CS introduces an anomaly in the holographic theory, and then makes a gauged Wess--Zumino--Witten term to appear in the holographic Goldstone bosons action. We consider a rather general case, but we later specify to QCD where a CS is needed to mimic the Adler--Bardeen chiral anomaly. Such term, indeed, automatically appears in the stringy version of holographic QCD discussed in \cite{Sakai}.

Our conclusions are in sect.~6; appendix~A contains a more formal derivation, which we will need in sect.~5, of the gauge--fixed holographic partition function. In appendix~\ref{appBoundaryTerms} we discuss how localized terms can be easily handled within the holographic procedure. Details on the calculation of the holographic effective action in the particular cases of flat and $AdS_5$ spaces are reported in appendix~\ref{appSolEOM}.

\section{Gauge Theories on the Interval: Holographic Approach}

In this paper we will be dealing with five--dimensional models compactified on a generic warped segment. The $5D$ metric is
\be
ds^2=a(z)^2\left(dx_\mu dx^\mu-\,dz^2\right)\,,
\label{metric}
\ee
where the fifth coordinate $z$ is defined on the interval $[z_{\UV},z_{\IR}]$, the warp function $a(z)$ is assumed to be regular and positive and the $4D$ coordinates have been rescaled to fix $a(z_{\UV})=1$. In our notations $4D$ indices are raised and lowered with the Minkowski metric $\eta={\textrm{diag}}(+,-,-,-)$. From the above general formula flat space is recovered for $a(z)=1$ while the RS set--up ({\it{i.e.}} $AdS_5$) corresponds to $a(z)=L/z$, $z_{\UV}=L$. 

As discussed in the Introduction, a powerful tool to deal with field theories on the warped space of eq.~(\ref{metric}) is provided by the holographic method. On general grounds, it consists in separating the $UV$ boundary value of the fields from their bulk fluctuations, treating them as distinct degrees of freedom. Of course, it is useful to take this approach when, after having separated the brane from the bulk, one integrates out the latter in order to obtain an effective theory for the holographic fields. In this section we apply the first step of the holographic program to the case of a gauge theory. The second step, {\it{i.e.}} the integration of the bulk degrees of freedom, will be taken in the next section.

Gauge theories in the generic warped space of eq.~(\ref{metric}) are our target. To introduce basic concepts of the holographic approach avoiding any technical complication, however, we find it useful to start with the simpler case of a massless scalar field in flat space.\footnote{Restricting to flat space we will avoid the complication due to the localized mass--terms which are commonly introduced in order to obtain a zero--mode. The general treatment of the scalars can be found in the appendices~B and C.} Having worked out this example, it will be simple to discuss gauge fields.

\subsubsection*{The Example of a Scalar}

Our starting point is the standard $5D$ action
\be
S\left[\Phi\right]\,=\,\frac12\int_{z_{\UV}}^{z_{\IR}}dz\left[\de_\mu\Phi\de^\mu\Phi\,-\,\left(\de_z\Phi\right)^2\right]\,,
\label{acts}
\ee
where we omitted the $d^4x$ integral and $\de_z$ is the derivative along the extra coordinate. The theory is defined by the partition function
\be
{\mc Z}\,=\,\int \mathcal{D}\Phi(x,z)_{b.c.}\,{\ds{e^{iS\left[\Phi\right]}}}\,,
\label{partfs}
\ee
where the ``$b.c.$'' subscript means that suitable boundary conditions are imposed on the field configurations one integrates over. The allowed boundary conditions are extracted by varying the action (\ref{acts})
\bea
\delta S\left[\Phi\right]\,&=&\,\int_{z_{\UV}}^{z_{\IR}}dz
\left[-\delta\Phi\de_\mu\de^\mu\Phi\,+\,\delta\Phi\de_z^2 \Phi\right]\nn\\&&
-\,\left[\delta \Phi\de_z\Phi\right]_{z_{\IR}}\,+\,\left[\delta \Phi\de_z\Phi\right]_{z_{\UV}}\,,
\label{actvar}
\eea
and imposing the boundary terms to vanish, so that the tree--level equation of motion coincides with the $5D$ Klein--Gordon equation. The two well known possibilities are Neumann ($\de_z\Phi=0$) or Dirichlet ($\Phi=0$) boundary conditions on each brane.

Once a particular set of conditions has been chosen, the standard KK approach consists in expanding $\Phi$ on a basis for functions on the segment which obey such conditions. Those functions are a numerable set and the $5D$ functional integral in eq.~(\ref{partfs}) is rewritten as an infinite product of $4D$ integrals. To discuss the holographic approach,
we must first of all make a distinction between the two boundaries and treat them in a different way. We will consider a field which is, for the moment, unconstrained at the $UV$ and satisfies a given boundary condition at the $IR$. We introduce a $4D$ scalar ``source'' field $\phi(x)$ and an ``holographic'' path--integral, in which the allowed $5D$ field configurations are only those which reduce to $\phi(x)$ at the $UV$ brane. The holographic partition function and the corresponding holographic action are defined as
\be
Z\left[\phi(x)\right]\,\equiv\,\ds{e^{iS_{h}\left[\phi(x)\right]}}\,=\,\int \mathcal{D}\Phi(x,z)_{\widehat{\Phi}(x)=\phi(x)}\,{\ds{e^{iS\left[\Phi\right]}}}\,,
\label{partfsc}
\ee
where $\widehat{\Phi}(x)\equiv\Phi(x,z_{\UV})$ and the $IR$ boundary conditions are understood. It is very easy to make contact with the standard definition (\ref{partfs}) of the partition function which also includes boundary conditions at the $UV$. Depending on whether $\Phi$ is Dirichlet or Neumann we have, respectively
\be
{\mc Z}\,=\,Z\left[\phi(x)=0\right]\,,\;\;\;\;\;\textrm{or}\;\;\;\;\;{\mc Z}\,=\,\int \mathcal{D}\phi(x)\,Z\left[\phi(x)\right]\,.
\label{equi}
\ee
In words, this is just the usual statement that in the Dirichlet case $\phi$ is a non--dynamical source to be put to zero at the end of the calculation, while Neumann boundary conditions are obtained by making the source dynamical. While the first (Dirichlet) case of eq.~(\ref{equi}) is trivial, one should be careful with the second one. Integrating eq.~(\ref{partfsc}) in $\mathcal{D}\phi$ as done in eq.~(\ref{equi}) is the same as integrating over the $5D$ field $\Phi$ without any boundary condition at the $UV$. Its variation $\delta\Phi$ is then also unconstrained and hence, to cancel the action variation (\ref{actvar}), the Neumann boundary condition $\left[\de_z\Phi\right]_{z_{\UV}}=0$ arises as an equation of motion. Notice that, even though the equivalence (\ref{equi}) has been written in a fully quantum form, we have only proven it at tree--level having shown that the classical equations of motion (including the boundary conditions) are the same.\footnote{In the
following section we will use holography to compute the $1$--loop Higgs effective potential
and the result matches the one obtained with the standard KK approach. This success, however,
does not rely on the validity of the equivalence at one--loop order because the effective
potential is completely determined by the tree--level spectrum.}

\subsubsection*{Holography for the Gauge Fields}

Let us now consider a $5D$ gauge theory with gauge group $G$ broken to $H\subset G$ at the $IR$. We take $G$ to be a compact Lie group and call $t^A=\{t^a,t^{\widehat{a}}\}$ its generators normalized to $2$Tr$[t^A\,t^B]=\delta^{AB}$.  The $t^a$s are the generators of the subgroup $H$ while the $t^{\widehat{a}}$'s generate the right coset $G/H$, in the sense that any element $g\in G$ can be uniquely decomposed as:
\be
g=\ds{ e^{i\alpha^A t^A}}=
\ds{
e^{i\gamma_{R}^{\widehat{a}} t^{\widehat{a}}}
\circ
e^{i h_{R}^a t^a} 
}\equiv \gamma_R[g]\circ h_R[g]\,,
\label{coset}
\ee
where $h_R[g]\in H$. With the metric (\ref{metric}) the gauge action reads
\be
S\,=\,\frac{1}{g_5^2}\int_{z_{\UV}}^{z_{\IR}}dz\frac{a(z)}2{\textrm{Tr}}\left[-F_{\mu\nu}F^{\mu\nu}\,+\,2 F_{\mu z}F^{\mu}_{\; z}\right]\,,
\label{gfa}
\ee
where the gauge field strength is $F_{MN}=F_{MN}^At^A=\partial_M A_N-\partial_N A_M-i[A_M,\,A_N]$ with $M=\{\mu,z\}$ and $A_{M}=A_{M}^At^A$.

As for the scalar, let us first of all determine the allowed boundary conditions by varying the action. We get
\bea
\delta S&=&\frac{2}{g_5^2}\int_{z_{\UV}}^{z_{\IR}}dz\,{\textrm{Tr}}\left[\delta A^\mu\left(a\,D^\nu F_{\nu\mu}+D_z(a\,F_{\mu z})\right)\,-\,\delta A_z\left(a\,D^\mu F_{\mu z}\right)\right]\nn\\
&&-\frac{2}{g_5^2}{\textrm{Tr}}\left[a\,F_{\mu z}\delta A^\mu\right]_{z_{\IR}}\,+\,\frac{2}{g_5^2}{\textrm{Tr}}\left[a\,F_{\mu z}\delta A^\mu\right]_{z_{\UV}}\,,
\label{gfvar}
\eea
where we defined $D_M \equiv \partial_M - i [A_M, \cdot]$.
To cancel the boundary variations and obtain the non--Abelian Maxwell equation in the bulk two possibilities are given. At each boundary and for each component $A^A$ of the gauge field either we take Dirichlet ($A_{\mu}^A=0$) or Neumann ($F_{\mu z}^A=0$) boundary conditions. The first choice induces a breaking of the transformations generated by $t^A$, while the symmetry is unbroken in the second case.\footnote{Dirichlet boundary conditions for $A_\mu$ are usually accompanied by Neumann conditions $\de_z A_z=0$ for $A_z$ while the usual form of the Neumann condition is $\de_z A_\mu=A_z=0$. The need for additional conditions is due to the commonly used gauge--fixing.} We have to choose the boundary conditions in such a way that the Neumann generators at each brane form a subgroup of $G$. At the $IR$ we have
\be
\left(F\right)_{\mu\,z}^a(x,z_{\IR})=0\,,\;\;\;\;\;\;\left(A\right)_{\mu}^{\widehat{a}}(x,z_{\IR})=0\,,
\label{bc}
\ee
while the $UV$ boundary conditions need not to be specified, at the moment.

Analogously to eq.~(\ref{partfsc}), the holographic action is defined as
\be
Z\left[B_\mu\right]\,\equiv\,\ds{e^{iS_{h}\left[B_\mu\right]}}\,=\, \int \mathcal{D}A_\mu(x,z)_{\widehat{A}_\mu=B_\mu}\mathcal{D}A_z(x,z)\exp{\left[iS\left[A\right]\right]}\,,
\label{partf}
\ee
where $\widehat{A}_\mu=A_\mu(x,z_{\UV})$ and we introduced $4D$ vector sources $B_{\mu}^A$ for each generator of $G$. As for the scalar, any boundary condition at the $UV$ can be implemented once we have the holographic action (\ref{partf}). If a given component $A_{\mu}^A$ is Dirichlet we simply have to put the corresponding source $B_{\mu}^A$ to zero, if it is Neumann we have to make it dynamical. If $\left[\delta A_{\mu}^A\right]_{z_{\UV}}\neq0$, indeed, the Neumann condition $\left[F_{\mu z}^A\right]_{z_{\UV}}=0$ arises as an equation of motion in order to cancel the action variation (\ref{gfvar}). To recover a standard $5D$ theory with $G$ broken to $H'\subset G$ at the $UV$, then, we just have to make dynamical the sources associated to $H'$ and put the others to zero.

By construction, the action is invariant under local $5D$ transformations $g(x,z)\in G$ which act on the gauge fields as
\be
A_M\,\rightarrow\,A^{(g)}_M\equiv g\left[A_M\,+\,i\,\partial_M\right] g^\dag\,.
\ee
The $IR$ boundary conditions (\ref{bc}), however, are only $H$--invariant and hence the allowed transformations must reduce to $H$ at the $IR$, {\it{i.e.}} $g(x,z_{\IR})=h\in H$. The action and the measure (including the $UV$ constraint $\widehat{A}=B$) in eq.~(\ref{partf}) are invariant under the ``bulk'' local group $\mathcal{G}_B$, which we define as the set of allowed transformations which reduce to the identity at the $UV$ brane, $\mathcal{G}_B\equiv\left\{g(x,z)\in G\,:\,g(x,z_{\IR})\in H\,,g(x,z_{\UV})=\I\right\}$. Due to $\mathcal{G}_B$ invariance, the integrand in eq.~(\ref{partf}) has flat directions and the path--integral is ill--defined. This requires a gauge--fixing of the local $\mathcal{G}_B$ group which we will discuss in the next subsection.

It is important to remark that the holographic action defined by eq.~(\ref{partf}) is gauge invariant under the full $4D$ local group $G$ in the sense that
\be
Z\left[B_{\mu}^{(\widehat{g})}\right]=Z\left[B_\mu\right]\,,
\label{ginv}
\ee
where $\widehat{g}(x)$ is a generic $4D$ local $G$ transformation. This is easily shown by performing a $5D$ gauge transformation $g(x,z)$ which reduces to $\widehat{g}(x)$ at the $UV$, $g(x,z_{\UV})=\widehat{g}(x)$, and of course belongs to $H$, $g(x,z_{\IR})=h(x)$, at the $IR$.\footnote{If seen as a map to $G/H$, $h(x)$ is just the identity. After rotating to the Euclidean, the problem of finding the $5D$ transformation $g$ which permits to prove eq.~(\ref{ginv}) is exactly the same of finding an homotopy which deforms into the identity the generic $S^4\rightarrow G/H$ map $\widehat{g}(x)$. We will assume that $\pi_4(G/H)$ is trivial, so that $g$ always exists.} When an $AdS/CFT$ interpretation is possible (as for $a(z)=L/z$ in eq.~(\ref{metric})), the $5D$ model would be dual to a $4D$ strongly coupled theory with global $G$ invariance and the holographic partition function (\ref{partf}) would be interpreted as the partition function of the $4D$ theory in the presence of sources
$B_{\mu}^A$ for the global currents $J_{\mu}^A$. Of course, the gauge invariance (\ref{ginv})
is necessary for this interpretation to be possible.

\subsubsection*{An ``Holographic'' Gauge--Fixing}\label{holgaugefix}

To fix the $\mathcal{G}_B$ gauge invariance of eq.~(\ref{partf}) we will go to the axial gauge, in which the fifth component $A_z$ of the gauge field is put to zero. It is clear that any consistent gauge--fixing would lead to the same physical results and one could even not fix the gauge at all as done for instance in \cite{Hirn:2005nr}. Our choice, however, appears particularly convenient because the scalar degrees of freedom are directly parametrized by a $4D$ Goldstone boson matrix $\Sigma$ (see eq.~(\ref{zgf})) and the $G$ invariance of the holographic theory is explicit. Moreover, in the axial gauge there are no ghosts and this is useful in view of a possible $AdS/CFT$ interpretation.

Starting from a generic gauge field configuration a unique local transformation $\ov{g}$ exists which puts $A_z$ to zero, {\it i.e.} $A^{(\ov{g})}_z=0$, and reduces to the identity on the $UV$ brane. This is the Wilson Line
\be
\ov{g}(x,z)\,=\,\mathcal{W}\left(z_{\UV},z\,;\,A\right)\,\equiv\,\textrm{P}\left\{\exp\left[-
i\int_{z_{\UV}}^{z}dz'\,A_{z}^A(x,z')\,t^A\right]\right\}\,,
\label{gbar}
\ee
on a straight path going from the $UV$ brane to a generic point $z$. In general, $\ov{g}$ does not belong to the symmetry group ${\mc G}_B$ since it does not reduce to $H$ at the $IR$ and therefore it cannot be used to reach the axial gauge. This was to be  expected, of course,
given that the theory has to contain physical scalars. In order for $\ov{g}$ to be useful we
have to formally restore the full $G$ invariance of eq.~(\ref{partf}) at the $IR$. To this end
we introduce a $4D$ field
\be
\Sigma(x)\,=\,\exp\left[i\,\sigma_{\widehat{a}}(x)t^{\widehat{a}}\right]\,,
\ee
and change the boundary conditions (\ref{bc}) by rotating them with $\Sigma^{-1}\in G$:
\be
\left(F^{\left(\Sigma^{-1}\right)}\right)_{\mu\,z}^a(x,z_{\IR})=0\,,\;\;\;\;\;\;\left(A^{\left(\Sigma^{-1}\right)}\right)_{\mu}^{\widehat{a}}(x,z_{\IR})=0\,.
\label{bcn}
\ee
Under $IR$ gauge transformations we take $\Sigma$ to transform as a Goldstone field
\be
\Sigma(x)\,\rightarrow\,\Sigma^{(g)}(x)\,=\,\gamma_R\left[g(x,z_{\IR})\circ\Sigma(x)\right]\,,
\label{goltr}
\ee
where $\gamma_R[.]$ is the projector from $G$ to the right $G/H$ coset defined by eq.~(\ref{coset}). It is easily checked that the boundary conditions (\ref{bcn}), given the transformation rule (\ref{goltr}) for the Goldstone boson field, are now invariant under generic transformations $g(x,z_{\IR})\in G$.

It should be noted that no extra dynamics has been introduced in the theory when we added the
Goldstone field $\Sigma$. We apparently added $dim(G/H)$ new real $4D$ (axion--like) scalars
$\sigma_{\widehat{a}}$, but we simultaneously enlarged the $IR$ gauge group by adding
$dim(G/H)$ symmetry transformations. We could put $\Sigma$ to $\I$ fixing in this way the
$G/H$ invariance at the $IR$ and recover the original theory with no $\Sigma$ field, boundary
conditions as in eq.~(\ref{bc}) and restricted gauge invariance. It is worth remarking that a
$4D$ Goldstone field localized at the $IR$ would naturally appear if, as usually assumed (see
for instance \cite{Csaki:2003dt, Barbieri:2003pr, Csaki:2003sh}), the symmetry--breaking
boundary conditions originate from some localized Higgs multiplet whose massive fluctuations
have been decoupled by taking the VEV to infinity. The present discussion could be more
directly applied to that case, in which the theory one starts with is fully $G$--invariant.
Our gauge--fixing procedure is however self--consistent even without this
interpretation, a more formal derivation can be found in appendix~A.

Instead of fixing the gauge group, up to now we have enlarged it. But now that $\ov{g}$ in eq.~(\ref{gbar}) belongs to the symmetry group we can easily go through a Fadeev-Popov procedure to fix the gauge. It is
\be
1\,=\,\int\mathcal{D}g(x,z)_{\widehat{g}=\I}\,\delta\left[A_{z}^{(g)}\right]\,\textrm{Det}\left\{D_z\left[A^{(g)}\right]\right\}\,,
\label{one}
\ee
where the Haar measure of $G$ is understood in the group integral. To prove the above identity one performs the change of variable $g\rightarrow g\circ \ov{g}$, so that in the new variable the delta function condition is simply $g=\I$ and the integral only receives contributions from the $g\sim \I$ region. For $g\sim \I$, $dg\sim\Pi_Ad\alpha_A$ and the identity is immediately demonstrated. The determinant in eq.~(\ref{one}) could be associated to a ghost action. However, since $A_{z}^{(g)}$ is the only component of $A^{(g)}$ which enters in $D_z[A^{(g)}]$ and it is fixed to zero by the delta function, the determinant is just a constant which we can drop. As customary, in the axial gauge there are no ghosts.

We now multiply our partition function by ``$1$'' written as in eq.~(\ref{one}) and with very standard manipulations we arrive to
\bea
&&\,Z^{g.f.}\left[B_\mu\right]\,=\,\iint\mc{D}\Sigma(x)\,\mc{D}A_{\mu}(x,z)_{\widehat{A}_\mu
=B_\mu}\exp\left[iS\left[A_\mu,A_z=0\right]\right]\,.
\label{zgf1}
\eea
For simplicity, in the above equation we did not write the $IR$ boundary conditions on the $A_\mu$ integral. Those are given by eq.~(\ref{bcn}) and depend on $\Sigma$. Note that the action $S\left[A_\mu,A_z=0\right]$ is still invariant under $G$ transformations which are constant along the extra coordinate. If we change variable $A_\mu\rightarrow A_{\mu}^{(\Sigma)}$ in the second functional integral, then, we can further simplify eq.~(\ref{zgf1}) by moving the dependence on $\Sigma$ from the $IR$ to the $UV$ brane. We finally get
\bea
&&\,Z^{g.f.}\left[B_\mu\right]\,\equiv\,\int\mc{D}\Sigma(x)\,\exp\left[iS_{h}\left[B_\mu,\,\Sigma\right]\right]\,\nn\\
&&=\,\iint\mc{D}\Sigma(x)\,\mc{D}A_{\mu}(x,z)_{\widehat{A}_\mu=B_{\mu}^{(\Sigma^{-1})}}
\exp\left[iS\left[A_\mu,A_z=0\right]\right]\,,
\label{zgf}
\eea
where the $IR$ boundary conditions are now given by eq.~(\ref{bc}) in which, since $A_z=0$, we can take $F_{\mu z}=-\de_z A_\mu$. In eq.~(\ref{zgf}) we also defined the holographic action $S_{h}\left[B_\mu,\,\Sigma\right]$. When a dual $AdS/CFT$ interpretation is possible this is the effective action for the Goldstone bosons in the presence of sources for the currents. 

From our procedure the gauge invariance of eq.~(\ref{zgf}) should be automatic. We can immediately check that $S_{h}\left[B_{\mu}^{(g)},\,\Sigma^{(g)}\right]=S_{h}\left[B_\mu,\,\Sigma\right]$. Indeed, due to eq.~(\ref{goltr}) $\Sigma^{(g)}=g\circ\Sigma\circ h$ so that
\be
\left(B^{(g)}\right)^{((\Sigma^{(g)})^{-1})}=B^{\left((\Sigma^{(g)})^{-1}\circ g\right)}=B^{\left(h^{-1}\circ\Sigma^{-1}\right)}=\left(B^{(\Sigma^{-1})}\right)^{(h^{-1})}\,.
\label{bountr}
\ee
Under a generic element of $G$, the boundary value $\widehat A_\mu=B_{\mu}^{(\Sigma^{-1})}$ of the $5D$ field only rotates with an $H$ transformation. The latter can be removed by a change of variable since both the action and the $IR$ boundary conditions are $H$ invariant. This demonstrates the gauge invariance of $S_{h}$ and, as a by-product, the gauge invariance of $Z^{g.f.}\left[B_\mu\right]$.

\section{The Tree-Level Holographic Action}\label{secEffAct}

In the previous section we have shown how to rewrite the partition function of a $5D$ gauge theory in terms of $4D$ holographic fields. With a bulk group $G$ broken to $H$ at the $IR$ and to $H'$ at the $UV$ the $4D$ dynamical degrees of freedom are the gauge fields $B'_\mu$ associated to $H'$ and the Goldstone bosons $\Sigma$ which parametrize the $G/H$ coset. The action $S_{h}[B',\Sigma]$ is defined in eq.~(\ref{zgf}), one simply has to put to zero the sources associated to $G/H'$. In this section we will compute at tree--level the quadratic effective holographic action for a generic $5D$ gauge theory with generic bulk fermion
content. Having in mind applications to models of Gauge--Higgs Unification, we will later use the action to compute the one--loop potential for $\Sigma$, {\it{i.e.}} for the Higgs. Details on the calculations can be found in appendix~C, in which we also include bulk scalars.

Let us start with the gauge fields, $S_{h}$ is defined by eq.~(\ref{zgf}) and to compute its quadratic part at tree--level one has to solve the linearized classical EOM's and then put the solutions back into the quadratic $5D$ action.\footnote{To obtain the full effective action one should solve the complete bulk equations of motion and this would require a suitable perturbative expansion \cite{Luty:2003vm}. This will be important in the next section in which we will use holography to compute interactions.} For what concerns the boundary conditions, one has to impose eq.~(\ref{bc}) at the $IR$ and the holographic condition $\widehat{A}=B^{(\Sigma^{-1})}$ at the $UV$. It is convenient to separate the longitudinal and transverse components of the gauge fields and to go to the $4D$ momentum space. We parametrize the solutions as
\be
\left\{
\begin{array}{l}
A^{\mu,A}_t (p, z) = \widehat A^{\mu,A}_t(p) f_t^A(p^2, z)\,,\\
A^{\mu,A}_l (p, z) = \widehat A^{\mu,A}_l(p) f_l^A(p^2, z)\,,
\end{array}
\right.\label{eqFieldExpansion}
\ee
where $\widehat A$ indicates as usual the value of the $5D$ field at the $UV$, so that $f_{t,l}^A(z_{\UV}) = 1$. The bulk EOM's are easily extracted by varying the Yang--Mills action in the axial gauge. This is done in appendix~C and the result is
\be\left\{
\begin{array}{l}
p^2 f_{t}^A
+ \displaystyle\frac{1}{a(z)} \partial_z (a(z) \partial_z) f_{t}^A = 0\,,\\
\displaystyle\frac{1}{a(z)} \partial_z (a(z) \partial_z) f_{l}^A = 0\,.
\end{array}
\right.\label{eqeqofmotapp1}
\ee
According to eq.~(\ref{bc}), the components associated to $H$ ({\it{i.e.}} $f_{t,l}^a$) are Neumann at the $IR$ while the $f_{t,l}^{\widehat a}$'s are Dirichlet. Given the $UV$ condition $f_{t,l}^A(z_{\UV}) = 1$ the solutions are completely determined, even though analytic expressions can only be obtained in very special cases.\footnote{Of course, the EOM for the longitudinal part is easily integrated.} Notice that $f_l^A(p^2, z)=f_t^A(p^2=0, z)$ and that the form of each $f^A$ only depends on whether the corresponding generators belong to $H$ or to $G/H$. We can then write
\be
\begin{array}{ll}
f_t^a(p^2, z)\,=\,F^+(p^2, z)\,,&\;\;\;\;\;f_t^{\widehat a}(p^2, z)\,=\,F^-(p^2, z)\,,\\
f_l^a(p^2, z)\,=\,F^+(0, z)\,,&\;\;\;\;\;f_l^{\widehat a}(p^2, z)\,=\,F^-(0, z)\,,
\end{array}
\label{eqscalwfsimpl}
\ee
where $F^{+(-)}$ is the solution with Neumann (Dirichlet) $IR$ boundary conditions. The explicit form of $F^{+(-)}$ in the cases of flat and $AdS_5$ space are reported in appendix~C.

It is straightforward to substitute the solutions back into the quadratic bulk action, one gets 
\be
S_{h} = - \frac{1}{2 g_5^2} \int d^4 x
\sum_A{\widehat A}^{A}_\mu\!
\left(\Pi^A_l P_l^{\mu\nu}
+ \Pi^A_t P_t^{\mu\nu}\right)\!
{\widehat A}^{A}_\nu
\,,
\label{eqholactgauge1}
\ee
where $\Pi^A_{t,l}(p^2) = \partial_z f_{t,l}^A(p^2,z_{\UV})$ and $P_{t,l}$ are the transverse and longitudinal projectors. The holographic fields $B_\mu$ and $\Sigma$ only enter the above equation through $\widehat A$. Remember that
\be\label{eqBCAmu}
{\widehat A}_\mu \,=\,{B'}_{\mu}^{(\Sigma^{-1})}\,=\, \Sigma^\dagger\left({B'}_\mu + i\,\partial_\mu\right) \Sigma\,,
\ee
where the only components of $B'$ are those associated to the $UV$ group $H'$. A particular but interesting case is when, as it happens for orbifold compactifications, a matrix ${\mc{P}}$ exists which commutes with all generators of $H$ and anticommutes with the others.
In this case eq.~(\ref{eqholactgauge1}) can be rewritten as
\bea
S_{h} &=& - \frac{1}{2g_5^2}\int d^4 x\ {\rm Tr}\left[{B'}_\mu^{(\Sigma^{-1})} \left(\Pi^0_l P_l^{\mu\nu}
+ \Pi^0_t P_t^{\mu\nu}\right) {B'}_\nu^{(\Sigma^{-1})}\right.\nonumber\\
&& \hspace{8em} \left.+ {\mc{P}} {B'}_\mu^{(\Sigma^{-1})} \left(\Pi^1_l P_l^{\mu\nu}
+ \Pi^1_t P_t^{\mu\nu}\right) {\mc P} {B'}_\nu^{(\Sigma^{-1})}\right]\,,\label{eqholactgauge2}
\eea
where $\Pi^{0,1}_{t}(p^2) = \partial_zF^+(p^2,z_{\UV}) \pm \de_zF^-(p^2,z_{\UV})$ and $\Pi^{0,1}_l =\Pi^{0,1}_t(0)$.

Let us now add massive bulk fermions $\Psi^I$ in some representation of the group $G$, their bulk action is in eq.~(\ref{eqFermionsAct}). The holographic procedure for the fermions has been discussed in \cite{Contino:2004vy}, we will briefly remind the subtleties which arise in this case. First of all, due to the fact that the bulk EOM's are of first order, the left-- and right--handed components can not be simultaneously used as effective degrees of freedom. Instead, for each component $\Psi^I$ one must choose either $\widehat\Psi_{L}^I$ or $\widehat\Psi_{R}^I$ as the holographic field. It is simpler to take the same chirality for all the components, we will use the left--handed part and then the $G$ multiplet $\chi_L=\widehat\Psi_{L}$ will be our holographic degree of freedom. Moreover, one must add to the action a localized mass--term, whose sign depends on the chirality of the holographic field. In our case we always have the same sign
\be\label{smass}
S_{m} = \frac{1}{2 g_5^2} \int_{UV} d^4 x \sum_I \left(\ov\Psi_L^I \Psi_R^I + h.c.\right)\,,
\ee
and the boundary term is $G$ invariant. To recover the original $UV$ boundary conditions, as usual, the Dirichlet sources must be put to zero while the Neumann ones must be made dynamical. It is correct to proceed in this way, the only problem is that the holographic theory would not contain any dynamical field associated to the components $\Psi^{i'}$ of the multiplet whose left--handed component is Dirichlet at the $UV$. The right--handed part of such fields might give rise to zero--modes whose dynamics could not be described in the holographic theory. As discussed in \cite{Contino:2004vy}, the solution consists in keeping $\widehat \Psi^{i'}_L$ dynamical and introduce right--handed Lagrange multipliers $\lambda_R^{i'}$ to dynamically impose the D boundary condition. One adds to the action the term
\be
S_{L.m.} = -\frac{1}{g_5^2} \int d^4x \sum_{i'} \left({\ov \chi}_L^{i'} \lambda_R^{i'} + h.c.\right)\,.
\label{eqLagrMult}
\ee
The Lagrange multipliers $\lambda_R^{i'}$ will describe the massless or light states associated to $\Psi^{i'}_R$. Notice that, if the action contains extra operators localized at the $UV$, the inclusion of the Lagrange multipliers is necessary. Such terms are  considered in appendix~\ref{appBoundaryTerms}, we will show how the holographic procedure
allows to treat them in a simple and straightforward way.

Having chosen the same chirality for all the sources, the full Lagrangian for $\Psi$ (including the boundary term (\ref{smass})) is gauge invariant and the manipulations of sect.~2 can easily be repeated. The fermion contribution to the gauge--fixed partition function (\ref{zgf}) can be schematically written as
\be
Z^{g.f.}_{f}\left[\chi_L,\Sigma\right]\equiv\exp{iS_{h}\left[\chi_L,\Sigma\right]}=\int\mc{D}\Psi(x,z)_{\widehat{\Psi}_L=\Sigma_f^{-1}\chi_L}\exp\left[iS_f\left[\Psi;A_\mu,A_z=0\right]\right]\,,
\label{zgffer}
\ee
where $\Sigma_f$ represents the Goldston boson matrix in the appropriate $G$ representation and $S_f$ is the full fermionic action which includes bulk and boundary terms (\ref{eqFermionsAct}, \ref{smass}) and possibly (\ref{eqLagrMult}). As before, we concentrate on the
quadratic part of the holographic action. The solutions of the EOM's (eq.~(\ref{eqfermEOM}))
can be parametrized as
\be
\Psi_L^I(p,z) = \widehat\Psi_L^I(p) f_L^I(p, z)\,,
\qquad \quad \Psi_R^I(p,z) = \frac{\pslash}{p} \widehat\Psi_L^I(p) f_R^I(p, z)\,,
\ee
with $f_L^I(p, z_{\UV}) = 1$ and suitable boundary conditions on the $IR$ boundary. As for the gauge field, the $f_{L,R}^I(p,z)$ functions have only two forms $f_{L,R}^I(p,z) = f_{L,R}^\pm (p,z)$ depending
on whether $\Psi^I_L$ satisfies N or D conditions at the $IR$ boundary.\footnote{The explicit form
of such functions for the $AdS_5$ and flat space cases can be found in appendix~C.}
Substituting into the action we find
\be\label{eqHolActFerm1}
S_{h} = \frac{1}{g_5^2} \int d^4 x \sum_I (\ov\chi_L \Sigma_f)^I\ \Pi_f^I(-\partial^2)\, i\,\dslash\, \left(\Sigma_f^\dagger \chi_L\right)^I
+ S_{L.m.}
\,
\ee
where $\Pi^I_f(p^2) = f_R^I(p, z_{\UV})/p$.
As for the gauge part, if the $IR$ boundary conditions are determined by the projection matrix ${\mc P}_f$ ({\it i.e.} $\Psi_{L,R}(z_{\IR}) = \pm {\cal P}_f \Psi_{L,R}(z_{\IR})$), the above equation can be written in the simpler form
\be\label{eqHolActFerm2}
S_{h} = \frac{1}{2g_5^2} \int d^4 x\ \ov\chi_L \Sigma_f\left(\Pi_f^0(-\partial^2)
+ \Pi_f^1(-\partial^2) {\mc P}_f\right)i\,\dslash\left(\Sigma_f^\dagger \chi_L\right)
+ S_{L.m.}
\,,
\ee
where $\Pi^{0,1}_f(p^2) = \left(f_R^+(p, z_{\UV}) \pm f_R^-(p, z_{\UV})\right)/p$. 

\subsection{The One-Loop Higgs Effective Potential}

The general formalism presented in the previous sections could be applied to several models of New Physics. For example, in the Higgsless models of ElectroWeak Symmetry Breaking (EWSB) \cite{Csaki:2003dt,Csaki:2003zu} one would consider a SM bulk gauge group $G=SU(2)\times U(1)_Y$ broken to $U(1)_{em}$ at the $IR$ and completely unbroken at the $UV$. In the holographic effective theory one would find the SM gauge bosons and the Goldstones which arise from the $IR$ breaking. The latter are eaten by the gauge fields becoming massive and the EWSB is obtained without introducing any physical scalar in the spectrum. The models of Gauge--Higgs Unification (or of Composite Higgs) \cite{Hatanaka:1998yp,Scrucca:2003ra,Panico:2006em,Contino:2003ve,Agashe:2004rs,Agashe:2005dk,Contino:2006qr} are more interesting for us, since they lead to physical $4D$ scalars whose
dynamics can be studied with our formalism. In this case one has an ``extended'' EW bulk gauge
group ($G=SU(3)$ in \cite{Scrucca:2003ra,Panico:2006em}, $G=SO(5)$ in \cite{Agashe:2004rs,Agashe:2005dk,Contino:2006qr}) broken
at the $IR$ in such a way that a complex Higgs doublet of scalars arises (respectively,
$H=SU(2)\times U(1)$ and $H=SO(4)$) as a Goldstone boson. The surviving $UV$ gauge group $H'$ is
the SM $SU(2)\times U(1)_Y$ and the EWSB comes from the Higgs taking a VEV. Since the $UV$
boundary conditions break part of the bulk group, the Higgs is really a pseudo--Goldstone
boson and these models are similar to the old Composite--Higgs scenario
\cite{GeorgiKaplan}.

The EWSB occurs radiatively in this scenario and it is then important to compute the one--loop Higgs potential. This is quite a difficult task with the standard KK approach (see for instance \cite{OdaFalkowski}) and the holographic procedure, as we will show below, is much simpler. Notice that the idea of computing the Higgs potential from the holographic action has already been used in \cite{Agashe:2004rs}. What we add is a precise recipe to derive the holographic Goldstone bosons action, which was obtained in \cite{Agashe:2004rs} by matching the $4D$ theory with the $5D$ results. In this section we will mostly keep the discussion general but we will later specify to a toy $SU(2)\rightarrow U(1)$ example in order to show how our final formulas can be applied to a concrete model.

At one loop, the effective potential is just the vacuum energy in the presence of a constant
background (which we denote by $\ov\Sigma$) for the Goldstone fields. In principle, we
should integrate at one--loop level the bulk degrees of freedom in eqs.~(\ref{zgf},
\ref{zgffer}), and later integrate the holographic fields. The Goldstones, however, only
appear in the $UV$ boundary conditions so that integrating out the bulk will not give any
contributions to the potential. The potential will only come from integrating over the
holographic fields, and at one--loop level the only part of their action which we need is the
quadratic tree--level one we already computed in the previous section.

As shown in section~2, the holographic effective action is gauge invariant. Once the non--dynamical sources are put to zero, the $H'$ invariance still survives and then a gauge--fixing is needed in order to compute the gauge field contribution to the effective potential.
The Landau gauge $\partial_\mu {B'}^\mu = 0$ is a particularly useful choice since no quadratic mixing can appear between the scalar fluctuations and ${B'}_\mu$. The quadratic action for ${B'}_\mu$ with a VEV for $\Sigma$ is (see eq.~(\ref{eqholactgauge1}))
\be
S_{h} = - \frac{1}{2 g_5^2} \int d^4 x
\sum_A\left[\overline\Sigma^\dagger {B'}_\mu \overline\Sigma\right]^A
\Pi^A_t P_t^{\mu\nu}
\left[\overline\Sigma^\dagger {B'}_\nu \overline\Sigma\right]^A\,,
\label{sgauge}
\ee
it can schematically be rewritten as
\be
S_{h} = - \frac{1}{g_5^2}\int d^4 x
\ P_t^{\mu\nu} \sum_{a',b'} {B'}_\mu^{a'}\, \Pi_g^{a',b'}(p^2,\overline\Sigma)\, {B'}_\nu^{b'}\,,
\ee
where $a'$ and $b'$ run over the generators of $H'$. Remembering that the ghosts do not contribute in the Landau gauge and that a transverse vector in $4D$ has $3$ real components, the gauge contribution to the Higgs potential reads
\be
V_g(\overline \Sigma) = \frac{3}{2} \int \frac{d^4 p_E}{(2 \pi)^4}
\log\left[{\rm Det}\left(\Pi_g(-p_{E}^2, \overline\Sigma)\right)\right]\,,
\label{eqeffpotgauge}
\ee
where we rotated the momenta to the Euclidean. It is easy to check that the Goldstone bosons fluctuations do not contribute to the potential, eq.~(\ref{eqeffpotgauge}) then contains the full $5D$ gauge contribution. Notice that the integrals involved in the effective potential are divergent, but they can always be made finite by subtracting the vacuum energy, {\it{i.e.}} the potential at $\ov\Sigma=0$. This is because the two solutions to the quadratic equations ($F^\pm$ in the case of the gauge field) become the same in the large Euclidean momentum limit. The inverse propagators $\Pi^A_t$ then become all equal so that the dependence on $\ov\Sigma$ drops from eq.~(\ref{sgauge}). The divergent part of the potential is therefore always independent of $\ov\Sigma$.

To compute the contribution of the fermions it is useful, as a first step, to integrate out the Lagrange multipliers
in the holographic action, {\it{i.e.}} to put to zero the sources associated to Dirichlet fields.\footnote{The same result would be obtained if including all the left--handed fermions and the Lagrange multipliers.} In this way one obtains the
quadratic action in $\chi_L$ from eq.~(\ref{eqHolActFerm1}):
\be
S_{h} = \frac{1}{g_5^2}\int d^4 x\ \sum_{i',j'} \ov\chi_L^{i'} \Pi_f^{i',j'}(\ov\Sigma) \chi_L^{j'}\,,
\ee
where $i'$ and $j'$ run over the left--handed fermion components with Neumann boundary conditions at the $UV$. The effective potential follows trivially:
\be
V_f(\ov \Sigma) = - 2 \int \frac{d^4p_E}{(2\pi)^4}
\log\left[{\rm Det}\left(\Pi_f(\ov\Sigma)\right)\right]\,.
\ee

\subsubsection{The $SU(2) \rightarrow U(1)$ Case}\label{subEffPotSU2}

The above formulas permit to derive the effective potential for a generic gauge group $G$ broken to arbitrary subgroups at the $UV$ and $IR$ boundaries. Applying them to a concrete model is extremely simple once the group--theoretical aspects have been worked out. To show this let us discuss explicitly the simple case of an $SU(2)$ bulk gauge theory broken to $U(1)$ (the same $U(1)$ subgroup) at both branes. This leads to a residual $U(1)$ invariance and to a charged physical Higgs scalar. We choose $\sigma^3/2$ to be the unbroken $SU(2)$ generator, where $\sigma^i$ are Pauli matrices. This pattern of symmetry breaking can be obtained by the projection matrix ${\mc P} = \sigma^3$. The Goldstone
boson matrix is
\be
\Sigma = \exp\left[i s_1 \frac{\sigma^1}{2} + i s_2 \frac{\sigma^2}{2}\right]
= \cos\left(\frac{s}{2}\right) \I
+ i \frac{s_i}{s}\sin \left(\frac{s}{2}\right)\sigma^i\,,
\ee
where $s_i(x)$ ($i = 1,2$) are the Goldstone boson fields and $s \equiv \sqrt{s_1^2+s_2^2}$. Using the unbroken $U(1)$ we align the Goldstone boson VEV along the $\sigma^2$ direction, thus we get
\be
\overline \Sigma {\mc P} \overline \Sigma^\dagger =
\left(
\begin{array}{cc}
\cos s & -\sin s\\
-\sin s & -\cos s
\end{array}
\right)\,.
\ee
At the $UV$, the only unbroken generator is $\sigma_3$ so that $B' = {B'}^3 \sigma^3/2$. From eq.~(\ref{eqholactgauge2}),
\be\label{eqEffActSU2}
S_{h} = - \frac{1}{4 g_5^2} \int d^4 x\, P_t^{\mu\nu}\
{B'}_\mu^3 \left(\Pi^0_t + \cos 2s\ \Pi^1_t\right) {B'}_\nu^3\,,
\ee
and the gauge contribution to the effective potential can be easily found to be
\be\label{eqEffPotSU2}
V_g (s) = \frac{3}{2} \int \frac{d^4 p_E}{(2 \pi)^4}
\log\left[\Pi^0_t + \cos 2s\ \Pi^1_t\right]\,.
\ee

Let us now take a bulk fermion $\Psi$ in the fundamental representation of $SU(2)$, with boundary conditions generated by the projection matrix ${\mc P}_f = \mc{P}$, {\it i.e.} $\Psi_{L,R} = \pm {\mc P} \Psi_{L,R}$ at both boundaries. Once we integrate out the Lagrange multiplier, the holographic action at the quadratic level becomes (see eq.~(\ref{eqHolActFerm2}))
\be
S_{h} = \frac{1}{2 g_5^2}\int d^4 x\ \ov\chi_L^1 (\Pi_f^0 + \cos s\, \Pi_f^1) i\,\dslash\,\chi_L^1\,,
\ee
and the fermion contribution to the effective potential is
\be
V_f(s) = - 2 \int \frac{d^4p_E}{(2\pi)^4}
\log\left[\Pi_f^0 + \cos s\ \Pi_f^1\right]\,.
\ee
The explicit expressions for $\Pi^{0,1}_{t,s,f}$ are reported in appendix~\ref{appSolEOM} for the $AdS$ and flat space cases, the result matches those obtained by KK computation in \cite{Scrucca:2003ra, OdaFalkowski}.

Notice that the results obtained for the gauge group $SU(2)$
can be used to derive the effective potential in more interesting
and complex models. When a single Higgs doublet is present, its VEV can be aligned along one of the broken generators and the system can be split in many $SU(2)$-like subsystems each with a different charge $q$ with respect to the Higgs VEV. In this case the effective potential can be computed summing the corresponding $SU(2)$ contributions with $s$ replaced by $q s$. Consider for instance $SU(3)$ broken to $SU(2) \times U(1)$ at both boundaries. In this case an $SU(2)$ doublet of Goldstone bosons arises. The gauge fields in the unbroken subgroup can be split into four $SU(2)$-like subsets with charges $q_1 = 1$, $q_2 = q_3 = 1/2$
and $q_4 = 0$. The gauge contribution to the effective potential is
\be
V^{SU(3)}_g(s) = V_g(s) + 2 V_g(s/2)\,.
\ee
Analogously we can treat fermion fields. For a bulk fermion in the fundamental
representation of $SU(3)$ we find a field with charge $q=1/2$ which gives a contribution
\be
V^{SU(3)}_{f,fund}(s) = V_f(s/2)\,,
\ee
while a bulk fermion in the symmetric representation gives a field with charge $q=1$
and a field with charge $q=1/2$ and in this case
\be
V^{SU(3)}_{f,sym}(s) = V_f(s) + V_f(s/2)\,.
\ee
These expressions for the effective potential coincide in the flat space case with the ones found in \cite{Scrucca:2003ra}.

\section{Holographic QCD}

Holographic QCD is a phenomenological attempt, inspired by $AdS/CFT$, of describing low energy (large $N_c$) strongly coupled QCD by means of a $5D$ weakly coupled model on the warped interval. One considers \cite{QCD1,QCD2} an $SU(3)_L\times SU(3)_R$ bulk gauge group which accounts for the global chiral symmetry of QCD and breaks it to its vector subgroup through the background profile of a bulk scalar field. An interesting limit is when the profile is exactly localized at the $IR$, in which case the breaking is equivalent to the one obtained by boundary conditions. Taking the limit does not significatively affect the degree of accordance of the model with real--world QCD, a detailed analysis of this case has been performed in \cite{Hirn:2005nr}. In this section we discuss this simplified limit of $AdS/QCD$ and then consider a $5D$ theory with chiral gauge group $G\,=\,\Gamma_L\times \Gamma_R$, with $\Gamma$ any compact Lie group. More explicitly,
each element $g$ of $G$ is
the direct product of two elements ($g_L$ and $g_R$) of $\Gamma$, {\it{i.e.}} $g=g_L\times
g_R$. For applications to QCD one takes $\Gamma=SU(2)$ or $SU(3)$ and interprets the
holographic partition function in eq.~(\ref{partf}) as the QCD partition function in the
presence of sources for the chiral group currents. The $IR$ boundary conditions are chosen to
break $G$ to its vector subgroup $\Gamma_V\sim \Gamma$ whose elements are couples $g=g_V\times g_V$
of equal elements $g_V\in\Gamma$ acting on the two chiral subspaces. The results of sect.~2
could be directly applied to this case, but this would lead us to a non--standard
parametrization of the Goldstone degrees of freedom. It is easier to restart from the
beginning and adapt the discussion of sect.~2 to the case of a chiral gauge group.

We have two $5D$ gauge fields $A_L$ and $A_R$, the first transforming under $g_L$ and the second under $g_R$. The $IR$ boundary conditions are
\be
A_{L,\mu}\,=\,A_{R,\mu}\,,\;\;\;\;\; F_{L,\mu z}\,=\,-F_{R,\mu z}\,.
\label{bcchiral}
\ee
By looking at eq.~(\ref{gfvar}) we see that eq.~(\ref{bcchiral}) ensures the cancellation of the boundary variations of the gauge action. Moreover, eq.~(\ref{bcchiral}) is covariant under the vector subgroup only, so that it restricts the allowed $5D$ local transformations to those which are vector--like at the $IR$. As in sect.~\ref{holgaugefix}, we want to enlarge the $5D$ group and allow any transformation at the $IR$. This is easily done by defining, as customary in QCD, a matrix
\be
U(x)\,=\,\exp\left[i\,\sigma_a(x) t^a\right]\,\in\Gamma\,,
\ee
which transforms under $G$ as:
\be
U\,\rightarrow\,U^{(g)}\,=\,g_R\circ U\circ g_{L}^{-1}\,.
\ee
The new boundary conditions
\be
\left(A_{L}^{(U)}\right)_\mu\,=\,A_{R,\mu}\,,\;\;\;\;\; \left(F_{L}^{(U)}\right)_{\mu z}\,=\,-F_{R,\mu z}\,,
\label{bcchiralnew}
\ee
are now covariant under the full chiral group. Note that $U$ plays exactly the same role as $\Sigma$ in sect.~\ref{holgaugefix}, using one or the other is just a reparametrization of the Goldstone degrees of freedom.

One can now go through the same manipulations of sect.~\ref{holgaugefix} and get
\bea
&&\hspace{-2em}Z^{g.f.}\left[l_\mu,\,r_\mu\right]\equiv\int\mc{D}U(x)\,\exp\left[iS_{\chi{\bf PT}}\left[l_\mu,\,r_\mu,\,U\right]\right]\,\nn\\
&&\hspace{-2em}\!\!=\iint\mc{D}U(x)\,\mc{D}A_{L,\mu}(x,z)_{\widehat{A}_{L,\mu}
=l_{\mu}^{(U)}}\,\mc{D}A_{R,\mu}(x,z)_{\widehat{A}_{R,\mu}=r_\mu}
\exp\left[iS\left[A_\mu,A_z=0\right]\right]\,,
\label{zgfchiral}
\eea
where the $IR$ conditions are simply given by eq.~(\ref{bcchiral}). One can easily check that eq.~(\ref{zgfchiral}) is gauge invariant under the full chiral group. The holographic action $S_{\chi {\bf PT}}\left[l_\mu,r_\mu,U\right]$ depends on the Goldstone matrix $U$ and on the chiral sources $l_\mu$, $r_\mu$. In $AdS/QCD$ this is interpreted as the action of chiral perturbation theory.

\subsection{The $\chi${\bf PT} Lagrangian at ${\mc O}(p^4)$ from Holographic QCD}

At tree--level, the $\chi${\bf PT} effective action is
\bea
S_{\chi{\bf PT}} &=& \frac{1}{g_5^2} \int d^4 x \int_{z_{\UV}}^{z_{\IR}} dz\, a(z)\ {\rm Tr}
\bigg[-\frac{1}{2} F_{L,\mu\nu} F_L^{\mu\nu} -\frac{1}{2} F_{R,\mu\nu} F_R^{\mu\nu}\nonumber\\
&& \hspace{11em}+ \partial_z A_{L,\mu} \partial_z A_L^{\mu}+ \partial_z A_{R,\mu} \partial_z A_R^{\mu}\bigg]\,,
\label{eqholQCDaction}
\eea
where $A_{L,R}$ satisfy the bulk EOM's with the $IR$ boundary conditions given
in eq.~(\ref{bcchiral}) and $UV$ boundary conditions
\be
\left\{
\begin{array}{l}
\widehat A_{L,\mu} = U\left(l_\mu + i \partial_\mu\right)U^\dagger\,,\\
\widehat A_{R,\mu} = r_\mu\,.
\end{array}
\right.\label{eqBCQCD1}
\ee
It is useful to define the vector and axial combinations of the gauge fields
\be
{\cal V}_\mu \equiv \frac{1}{2}(A_{L,\mu} + A_{R,\mu})\,,
\qquad\qquad
{\cal A}_\mu \equiv \frac{1}{2}(A_{L,\mu} - A_{R,\mu})\,,
\label{eqdefAV}
\ee
which satisfy simple boundary conditions at the $IR$
\be
\left\{
\begin{array}{l}
\partial_z {\cal V}_\mu (x, z_{\IR}) = 0\,,\\
{\cal A}_\mu (x, z_{\IR}) = 0\,.
\end{array}
\right.
\ee

Of course, computing $S_{\chi{\bf PT}}$ from eq.~(\ref{eqholQCDaction}) would require to solve the full bulk EOM's which include all the interactions, and this can only be done order by order in a given perturbative expansion. As customary in $\chi{\bf PT}$, we expand in powers of the momentum and treat the external sources $l_\mu$ and $r_\mu$ as terms of ${\mc O}(p)$. Given that $U$ is of ${\mc O}(p^0)$, eq.~(\ref{eqBCQCD1}) implies that the boundary value of the fields (${\widehat A}_{L,R}$, $\widehat{\cal V}$ and $\widehat{\cal A}$) are also of ${\mc O}(p)$. Our present goal is to determine the holographic effective action up
to ${\mc O}(p^4)$, which was first obtained in \cite{QCD2} (for the simplified model we are considering see \cite{Hirn:2005nr}) after a long KK calculation. In the mixed momentum--space representation we can expand the solutions of the full EOM's as
\be
\left\{
\begin{array}{l}
{\cal V}_\mu (p, z) = {f^0_{V}}(z)\, \widehat{\cal V}_\mu (p)
+{\cal V}_{\mu}^{(3)}(p,z,\widehat{\cal V},\widehat{\cal A}) \,,\\
{\cal A}_\mu (p, z) = {f^0_{A}}(z)\, \widehat{\cal A}_\mu (p)
+{\cal A}_{\mu}^{(3)}(p,z,\widehat{\cal V},\widehat{\cal A}) \,,
\end{array}
\right.
\label{eqQCDfieldexp}
\ee
where ${\cal V}_{\mu}^{(3)}$ and ${\cal A}_{\mu}^{(3)}$, whose explicit form we will not need, represent ${\mc O}(p^3)$ contributions to the solutions. The latter terms are trilinear combinations of the momentum $p_\mu$ and of the boundary fields $\widehat{\cal V}_\mu (p)$ and $\widehat{\cal A}_\mu (p)$. It is important to remark that the tensorial structure of the solutions implies that no ${\mc O}(p^2)$ terms appear in eq.~(\ref{eqQCDfieldexp}) and that the next correction will be of ${\mc O}(p^5)$. The first term of each expansion is a solution of the linearized bulk EOM's (eq.~(\ref{eqeqofmotapp1})) at zero momentum. At the $UV$, ${f^0_{V,A}} (z_{\UV}) = 1$ and they are, respectively, Neumann or Dirichlet at the $IR$ boundary. The higher order terms, on the contrary, vanish at the $UV$ while their $IR$ boundary conditions are the same as the corresponding zero--order terms.

Substituting the solutions of the EOM's into the effective action, we see that the terms in the first line of eq.~(\ref{eqholQCDaction}) only contribute to the ${\mc O}(p^4)$, and just the leading order of the series in eq.~(\ref{eqQCDfieldexp}) needs to be plugged in.
On the other hand, the terms in the second row of eq.~(\ref{eqholQCDaction}) will give a ${\mc O}(p^2)$ term when substituting the leading (${\mc O}(p)$) terms of eq.~(\ref{eqQCDfieldexp}) while an ${\mc O}(p^4)$ term could arise when taking one ${\mc O}(p)$ and
one ${\mc O}(p^3)$ term. The latter contribution, however, vanishes. This is easily verified by integrating by parts $\de_z$ and remembering that the ${\mc O}(p^3)$ terms vanish at the $UV$ boundary while the ones of ${\mc O}(p)$ verify the linearized EOM's at zero momentum. Thus, the effective action at ${\mc O}(p^4)$ can be written as
\bea
S_{h} &=& -\frac{2}{g_5^2} \int_{UV} d^4x\ {\rm Tr}
\left[{\cal V}_\mu \partial_z {\cal V}^\mu + {\cal A}_\mu \partial_z {\cal A}^\mu\right]\nonumber\\
&&-\frac{1}{2 g_5^2} \int d^4 x \int_{z_{\UV}}^{z_{\IR}} dz
\ a(z) {\rm Tr}\left[F_{L,\mu\nu}F_L^{\mu\nu} + F_{R,\mu\nu}F_R^{\mu\nu}\right]\,,
\label{eqeffactexp}
\eea
where the fields are now simply the solutions of the linearized bulk EOM's at zero momentum. Notice that in this
expression the first line gives ${\mc O}(p^2)$ terms, while ${\mc O}(p^4)$ operators come from the second one.

The linearized bulk EOM's (\ref{eqeqofmotapp1}) at zero momentum can be
analytically solved
\be
\left\{
\begin{array}{l}
{f^0_{V}}(z) = 1\,,\\
{f^0_{A}}(z) = \displaystyle \left(\int_z^{z_{\IR}} \frac{dz'}{a(z')}\right)
\left(\int_{z_{\UV}}^{z_{\IR}} \frac{dz'}{a(z')}\right)^{-1}\,,
\end{array}
\right.\label{eqWavefunctSimpl}
\ee
and it follows from eq.~(\ref{eqBCQCD1}) that
\be
{\widehat{\cal A}}_\mu = \frac{i}{2} U (D_\mu U)^\dagger = -\frac{i}{2} (D_\mu U) U^\dagger
\equiv u_\mu\,,
\ee
where we defined $D_\mu U = \partial_\mu U + i U l_\mu - i r_\mu U$. The kinetic term for the Goldstone bosons, {\it{i.e.}} the ${\mc O}(p^2)$ action, is immediately obtained from the first line of eq.~(\ref{eqeffactexp}).\footnote{Notice that $\partial_z f_V^0(z_{\IR}) = 0$,
and thus the term containing ${\cal V}_\mu$ in the first line of eq.~(\ref{eqeffactexp})
vanishes.} It is
\be
S_{h}^{(2)} = \frac{f_\pi^2}{2} \int d^4 x\ {\rm Tr}
\left[(D_\mu U) (D^\mu U)^\dagger\right]\,,
\ee
where the Goldstone boson decay constant $f_\pi$ is given by
\be
f_\pi =\displaystyle \frac{1}{g_5} \left(\int_{z_{\UV}}^{z_{\IR}} \frac{dz}{a(z)}\right)^{-1/2}\,.
\ee

Using the boundary conditions in eq.~(\ref{eqBCQCD1}), the definitions of ${\cal V}$ and
${\cal A}$ in eq.~(\ref{eqdefAV})
and the solutions of the bulk EOM's in eq.~(\ref{eqWavefunctSimpl}),
one gets
\bea
F_{L,\mu\nu} + F_{R,\mu\nu} &=& f_{+\mu\nu} + i\frac{1 - {f^0_A(z)}^2}{2} [u_\mu, u_\nu]\,,\\
F_{L,\mu\nu} - F_{R,\mu\nu} &=& f^0_A(z)\ f_{-\mu\nu}\,,
\eea
where we defined
\be
f_{\pm\mu\nu} \equiv U\, l_{\mu\nu}\, U^\dagger \pm r_{\mu\nu}\,,
\ee
with $l_{\mu\nu}$ and $r_{\mu\nu}$ the field strengths obtained from $l_\mu$ and $r_\mu$.
Substituting these expressions in the second line of eq.~(\ref{eqeffactexp}),
we find the ${\mc O}(p^4)$ terms of the effective action
\bea
S_{h}^{(4)} &=& -\frac{1}{4 g_5^2}\int d^4 x \int_{z_{\UV}}^{z_{\IR}} dz\ a(z)\,
{\rm Tr}\bigg\{-\frac{(1 - {f^0_A(z)}^2)^2}{2} u_\mu u_\nu \left[u^\mu, u^\nu\right]\nonumber\\
&& \hspace{5em} + 2 i (1-{f^0_A(z)}^2) f_{+\mu\nu} u^\mu u^\nu
+ f_{+\mu\nu} f_+^{\mu\nu} + {f^0_A(z)}^2 f_{-\mu\nu} f_-^{\mu\nu}\bigg\}\,.
\label{eqeffactQCD4}
\eea

In the $\Gamma = SU(3)$ case it is trivial to express the above action in the standard form of chiral perturbation theory \cite{Gasser:1984gg}.
From eq.~(\ref{eqeffactQCD4}) we find the values of the coefficients
\be
\left\{
\begin{array}{l}
L_1 = \displaystyle\frac{1}{16 g_5^2} \int_{z_{\UV}}^{z_{\IR}} dz\ a(z) (1-{f^0_A}^2)^2\\
\rule{0pt}{2em}L_{10} = \displaystyle-\frac{1}{2 g_5^2} \int_{z_{\UV}}^{z_{\IR}} dz\ a(z) (1-{f^0_A}^2)\\
\rule{0pt}{2em}H_1 = \displaystyle-\frac{1}{4 g_5^2} \int_{z_{\UV}}^{z_{\IR}} dz\ a(z) (1+{f^0_A}^2)
\end{array}
\right.
\,, \qquad\quad
\left\{
\begin{array}{l}
L_2 = 2 L_1\\
\rule{0pt}{1.5em}L_3 = -6 L_1\\
\rule{0pt}{1.5em}L_9 = -L_{10}
\end{array}
\right.
\,,
\ee
which are in agreement with those derived in \cite{Hirn:2005nr}.\footnote{The factor of $2$ difference with the result of \cite{Hirn:2005nr} is due to the normalization of the $SU(3)$ generators.}

\section{Holographic Anomaly}\label{secAnomaly}

It is interesting to study, using the holographic perspective which we
adopted in this paper, the consequences of adding a CS
term to the $5D$ gauge action. In this section we will work out the holographic Goldstone
bosons Lagrangian and find that, as expected \cite{Hill:2006wu}, the CS
makes a gauged Wess--Zumino--Witten (WZW) term \cite{Wess:1971yu,Witten:1983tw,Chu:1996fr}
appear. We will initially consider a quite general case, but we will
later specify to $AdS/QCD$, where a CS term is needed in order to mimic
the Adler--Bardeen chiral anomaly. Notice that, from a purely $5D$ point
of view, the CS is usually introduced
\cite{Anomalies} in order to cancel localized gauge 
anomalies \cite{Arkani-Hamed:2001is}. Our approach, however, is different.
We will simply add the CS to the action and study its consequences on the
holographic theory.
For simplicity, we will not consider bulk fermions at all, so that there
is clearly no localized anomaly
to be cancelled. In the presence of a bulk fermion content which gives rise to localized anomalies the analysis of this section should be generalized taking also into account the anomalous variation of the fermionic measure, and not only the one of the CS term as we will do here.

Let us consider, as in sect.~2, a pure gauge theory with bulk group $G$
broken to $H$ at the $IR$ and add to the gauge action the term
\be
S_{CS}\,=\,-i\, c\int \omega_5(A)\,=\,-i\, c\int {\rm Tr}\left[A (d A)^2
+ \frac{3}{2} A^3 (d A) + \frac{3}{5} A^5\right]\,,
\label{CS}
\ee
where $c$ is a real coefficient and we introduced a matrix form notation
defining $A=-iA_{M}^AT^A\,dx^M$.\footnote{In this section we use many definitions and results of \cite{Chu:1996fr}.} We want to explore
the consequences of including $S_{CS}$ in the definition (\ref{partf}) of
the holographic partition function. First of all we observe that, under a
generic infinitesimal gauge transformation
\be
\delta_\alpha S_{CS}\,=\,-i\,c\int
d\omega_4^1(\alpha,A)\,=\,i\,c\int\omega_4^1(\alpha(z_{\UV}),A(z_{\UV}))\,-\,i\,c\int\omega_4^1(\alpha(z_{\IR}),A(z_{\IR}))\,,
\label{cstr}
\ee
so that the gauge--invariance of the action has been spoiled. In
eq.~(\ref{partf}), however, the bulk gauge group ${\mc G}_B$ is gauged,
meaning that it is used to remove unphysical degrees of freedom and make
the theory consistent. For this reason, no anomalous variations of the
action under ${\mc G}_B$ can be allowed. Remembering that ${\mc G}_B$
elements reduce to the identity at the $UV$ and to the $H$ subgroup at the
$IR$, and using the $IR$ boundary conditions (\ref{bc}) for the gauge
fields, this translates into the condition
\be
{\textrm{Tr}}(T^a \{T^b,T^c\})\,=\,0\,,
\label{anfh}
\ee
on the $H$ generators which define the CS. The $T^a$'s, therefore, must
provide an anomaly--free representation (whose existence we are going to
assume in the following) of the $H$ subgroup. Eq.~(\ref{anfh}) ensures the ${\mc G}_B$ invariance
of the action as it makes the $IR$ term in eq.~(\ref{cstr}) vanish. The $UV$ term in the
variation (\ref{cstr}) is, on the contrary, perfectly allowed for what
concerns the definition of the holographic partition function. It simply
changes eq.~(\ref{ginv}) into
\be
Z\left[B_{\mu}^{(\widehat{g})}\right]\,=\,\exp\left[-\,c\int\omega_4^1(\widehat\alpha,B)\right]\,Z\left[B_{\mu}\right]\,,
\label{anomaly0}
\ee
where $\widehat{g}=\exp[i\widehat{\alpha}]$ is an infinitesimal gauge
transformation. If some of the sources have to be made dynamical, of
course, one should also worry about the cancellation of the $UV$ anomaly
in eq.~(\ref{anomaly0}). This could be done, for instance, by adding
localized fermions in a suitable representation.

When an $AdS/CFT$ interpretation is possible, $Z[B]$ is the partition
function of a $G$--invariant $4D$ theory in the presence of sources for
the currents and eq.~(\ref{anomaly0}) corresponds to a $4D$ anomaly. The
global group $G$ is spontaneously broken to $H$ and eq.~(\ref{anfh})
states that the unbroken group is anomaly free. This is very much the same
as in QCD, where the global symmetry $SU(3)_L\times SU(3)_R$ is spoiled by
the chiral anomaly while the unbroken vector subgroup is anomaly free.
Eq.~(\ref{anomaly0}), however, is not yet enough to mimic the standard QCD
Adler--Bardeen anomaly, since the latter preserves the vector invariance.
More generally, in a $4D$ theory with anomalous global symmetry group $G$ which is spontaneously broken to an anomaly free $H$, one adopts a regulator in which the $H$ invariance is preserved in
all correlators so that the anomaly $G_{\widehat\alpha}(B)$ vanishes, for a generic
gauge field configuration, whenever $\widehat{\alpha}\in Lie(H)$. The anomaly
$\int\omega_4^1(\widehat\alpha,B)$ which appears in eq.~(\ref{anomaly0}),
on the contrary, only vanishes (due to
eq.~(\ref{anfh})) when the sources also are restricted to $H$, {\it{i.e.}}
when $B,\,\widehat\alpha\in Lie(H)$.
In the language of Ref.~\cite{Chu:1996fr}, eq.~(\ref{anomaly0}) is the
``canonical'' anomaly
while what would be needed is the ``shifted'' anomaly. The two forms of
the anomaly correspond
to different choices of the regulator and one can convert one into the
other by adding
suitable local counterterms to the action. Following \cite{Chu:1996fr}, we
consider a
``shifted'' CS term
\be
{\widetilde S}_{CS}\,=\,-i\,c\int {\widetilde\omega}_5(A_h,A)\,
=\,-i\,c\int\left[\omega_5(A) \,+\,
dB_4(A_h,A)\right]\,=\,S_{CS}\,+\,i\,c\int_{UV} B_4(B_h, B)\,,
\label{cstl}
\ee
where $A_h$, $B_h$ are the restrictions of $A$, $B$ to $Lie(H)$ and $B_4$ is
defined by \cite{Chu:1996fr}
\be
B_4(A_h, A) = \frac{1}{2}{\rm Tr}\left[(A_h A - A A_h) (F + F'_h)
+ A A_h^3 - A_h A^3 + \frac{1}{2} A_h A A_h A\right]\,,
\label{eqB4}
\ee
with $F'_h = dA_h + A_h^2$ and $F = dA + A^2$.
In eq.~(\ref{cstl}) we used
the fact that, due to the boundary conditions, $A(z_{\IR})=A(z_{\IR})_h$
and $B_4(A_h, A_h)=0$. If ${\widetilde S}_{CS}$, and not ${S}_{CS}$, is added
to the gauge action, the anomalous variation of the holographic partition
function becomes
\be
Z\left[B_{\mu}^{(\widehat{g})}\right]\,=\,\exp\left[-c\,G_{\widehat\alpha}(B)\right]\,Z\left[B_{\mu}\right]\,,
\label{anomaly}
\ee
where
$G_{\widehat\alpha}=-\int\delta_{\widehat\alpha}{\widetilde{\omega}}_5$ is
the shifted anomaly as defined in \cite{Chu:1996fr}. It vanishes, for any $B\in Lie(G)$, when $\widehat\alpha\in Lie(H)$.

Having identified the correct term to be added to the $5D$ action, let us
now see how it affects the holographic effective action. It is well known
that the anomalous variation (\ref{anomaly}) of the partition function
{\emph{requires}} the presence of the gauged WZW term in
the effective action of the Goldstone bosons.
The latter term, indeed, is not gauge--invariant and it is precisely designed to reproduce
the anomaly (\ref{anomaly}). Showing that it arises, as
we will do in the following, is then just a check of internal consistency.
It is worth remarking that, from the purely $5D$ point of view, one could
also use the canonical CS (\ref{cstr}), the manipulations which follow
would not change significantly. Of course, having failed to reproduce the
anomaly (\ref{anomaly}), one would not obtain the correct gauged WZW. From
the point of view of holographic QCD, on the contrary, the shifted CS is the only correct
choice.

In order to fix the ${\mc G}_B$ gauge invariance of the holographic
partition function we cannot simply follow sect.~2. The CS, indeed, spoils
the invariance of the gauge action under generic $G$ transformations, so
that $G$ is broken to $H$ not only by the $IR$ boundary conditions
(\ref{bc}). The trick which we used in eq.~(\ref{bcn}) to restore the full
bulk $G$ invariance, therefore, is not useful in the present case. We will
need a more formal, but completely equivalent, gauge--fixing procedure, which is discussed in appendix~A. The final result is
\be
Z^{g.f.}\left[B_\mu\right]\,=\,\int{\mc D}\Sigma\int \mathcal{D}A_\mu(x,z)_{
\widehat{A}_\mu=B_{\mu}^{(\Sigma^{-1})}
}
\exp{\left[
iS\left[A^{(\Lambda)}\right]
\right]}\,,
\label{gff}
\ee
where $A_M = \{A_\mu, 0\}$ and
$\Lambda$ is any $5D$ transformation which interpolates from $\Sigma$
at the $UV$, ${\Lambda}(z_{\UV})=\Sigma$, to the identity at the $IR$,
${\Lambda}(z_{\IR})=\I$. When rotated to the Euclidean, $\Lambda$ is an
extension of $\Sigma$ from the $4$--sphere $S^4$ of space--time to a $5D$
disk $D_5$, with boundary $S^4$, obtained by shrinking the $IR$ brane to a
point.

The CS term ${\widetilde S}_{CS}$ gives the only non--trivial contribution
to $S\left[A^{(\Lambda)}\right]$. When the latter is absent the action is
invariant and eq.~(\ref{gff}) reduces to eq.~(\ref{zgf}). We define
\be
S_{wzw}\, =\,i\, c\int\left[\widetilde\omega_5(A_h,A) \, -\,
\widetilde\omega_5((A^{(\Lambda)})_h,A^{(\Lambda)}) \right]\,,
\label{eqwzw}
\ee
and rewrite
\be
S\left[A^{(\Lambda)}\right]\,=\,S_{wzw}\,-\,i\,c\int\widetilde\omega_5(A_h,A)\,+\,S_{g}\left[A\right]\,,
\label{rotact}
\ee
where $S_{g}$ is the standard gauge action. We added the term ${\widetilde
S}_{CS}=i\,c\int\widetilde\omega_5(A_h,A)$ to the definition of $S_{wzw}$
in order to make it vanish when $\Sigma=\I$. It should be noted that,
thanks to eq.~(\ref{bountr}), ${\widetilde S}_{CS}$ does not contribute to
the anomalous variation of the gauge--fixed partition function (\ref{gff})
since $\widetilde\omega_5(A_h,A)$ is $H$ invariant. One could check that,
on the contrary, $S_{wzw}$ varies and that its variation reproduces the
anomaly in eq.~(\ref{anomaly}).

It is convenient to rewrite $S_{wzw}$ in a more explicit form, this will
also allow us to check that it coincides with the result of
\cite{Chu:1996fr}. Using manipulations which are similar to those
explained before to eq.~(116) of \cite{Chu:1996fr}, and noticing that
$A^{(g)}$ as defined in that paper (eq.~109) corresponds to $A^{(g^{-1})}$ in our
conventions, we get
\bea
\widetilde\omega_5(A_h,A) \, -\,
\widetilde\omega_5((A^{(\Lambda)})_h,A^{(\Lambda)})
&&= \omega_5(\Lambda^{-1} d \Lambda)
\,-\,dB_4((A^{(\Lambda)})_h,A^{(\Lambda)})\nn\\
&&+\, dB_4((A)_h,A)\,+\, dB_4(- d \Lambda \Lambda^{-1}, A^{(\Lambda)})\,.
\label{eqmanWZW}
\eea
To obtain the WZW action of eq.~(\ref{eqwzw}) we must integrate
eq.~(\ref{eqmanWZW}) over the $5D$ space. Since $A(x,
z_{\IR}) \in Lie(H)$ and $\Lambda(z_{\IR})=\I$, no $IR$ contribution comes
from integrating the exact forms in the r.h.s. of eq.~(\ref{eqmanWZW}).
Using the $UV$ holographic condition $\widehat{A}=B^{(\Sigma^{-1})}$ and
remembering that $\Lambda(z_{\UV})=\Sigma$, we get
\bea
S_{wzw}\,=&&\,i\,c\int_5\omega_5(\Lambda^{-1} d \Lambda)\,+\,i\, c \int
B_4((B)_h,B) \nn\\
&& -i\, c
\int\left[B_4\left((B^{\Sigma^{-1}})_h,B^{\Sigma^{-1}}\right)+B_4(- d
\Sigma \Sigma^{-1}, B)\right]\,,
\label{WZWexpl}
\eea
where $\int_5$ represents the $5D$ space--time integral. The above formula
coincides with the result of \cite{Chu:1996fr} (eqs.~(115, 116)), given
that $\Lambda(z)$ is an extension of the Goldstone boson matrix $\Sigma$
to the disk $D_5$ obtained by shrinking the $IR$ boundary to a point. This
disk has however the opposite orientation than the one considered in
\cite{Chu:1996fr}, this explains the seeming sign difference.

Notice that $\omega_5(\Lambda^{-1} d \Lambda)$ is proportional to the
Maurer--Cartan $5$-form (see eq.~(\ref{CS})), so that its integral only
depends on the boundary value of $\Lambda$, {\it{i.e.}} on the holographic
field $\Sigma$.\footnote{If $\pi_5(G)$ is non--trivial, the integral can
also depend on the topology of $\Lambda$, in the sense that two
topologically inequivalent extensions of $\Sigma$ give different results.
The difference is however quantized, so that it does not affect the
path--integral if $c$, as we will show below, is also quantized.} We can
then rewrite the gauge--fixed holographic action as
\bea
&&\hspace{-3.5em}e^{iS_h[B_\mu,\Sigma]} = \exp\left[i S_{wzw}\left[B_\mu,
\Sigma\right]\right]\nn\\
&&\hspace{-1.em}\times \int {\mc D} A_\mu(x,z)_{\widehat A_\mu =
B_\mu^{(\Sigma^{-1})}} \exp\left[i S_{g}[A_\mu, A_z=0]
- i \widetilde S_{CS}[A_\mu, A_z=0]\right]\,,\label{eqshch}
\eea
where we used the fact that the WZW term depends only on the holographic
fields and factorized it out of the functional integral. As discussed
above, the $\widetilde S_{CS}[A]$ term is invariant under $H$
transformations, so that the functional integral term of
eq.~(\ref{eqshch}) is perfectly $G$--invariant. The above equation shows
that, in the presence of the CS term, the holographic action splits in two
parts. The first is simply $S_{wzw}$ and its gauge variation reproduces
the anomaly in eq.~(\ref{anomaly}), while the second one is
gauge--invariant.

In the above discussion we assumed (see Footnote~12) that, if $\pi_5(G)$
is non trivial, the coefficient $c$ of the CS term is quantized. We will
now prove that this assumption is necessary to ensure the consistency of
our $5D$ model. As we already discussed, the bulk local ${\mc G}_B$
invariance of the $5D$ action is gauged. Therefore, as we already did to
derive eq.~(\ref{anfh}), we must require $\exp[i{\widetilde S}_{CS}]$ to be
invariant under any ${\mc G}_B$ transformation. In particular, let us
consider a generic ${\mc G}_B$ element $g$ which reduces to the identity at
the $IR$ brane also.
After rotating to the Euclidean, such transformations are maps from $S^5$
(obtained by shrinking both boundaries to a point) to the group $G$. Let
us now start from the trivial $5D$ field configuration $A=0$ and vary the
CS term in eq.~(\ref{cstl}) with $g$, the variation is just the integral
of the CS form $\omega_5$ on the pure gauge configuration $A = 0^{(g)}$.
As we already mentioned, $\omega_5(gdg^{-1})=-\omega_5(g^{-1}dg)$ is
proportional to the Maurer--Cartan $5$-form and thus its integral on $S^5$
is quantized and depends on the $\pi_5(G)$ homotopy class to which the map
$g$ belongs. Summarizing, ${\mc G}_B$ invariance requires, for any $g\in
G$
\be
\frac{i\, c}{2\pi} \int_{S^5} \omega_5( g^{-1} d g)\, \in\, \ZZ\,.
\label{eqquantwzw}
\ee
If $\pi_5(G)=0$, the above equation gives no restriction on $c$ since
$\int_{S^5} \omega_5( g^{-1} d g)=0$. If, on the contrary, $\pi_5(G)$ is
non--trivial, $\int_{S^5} \omega_5( g^{-1} d g)=i\,\gamma\,n$ with $n$
integer and eq.~(\ref{eqquantwzw}) imposes $c$ to be quantized:
\be
c\,=\,\frac{2 \pi\, m}{\gamma}\,,\label{eqquantcond}
\ee
where $m$ is an integer.

In the framework of $4D$ effective action, the quantization condition of
the WZW term \cite{Witten:1983tw, D'Hoker:1994ti} seems weaker than what
we find here. It depends indeed on the homotopy of $G/H$, and not of $G$.
It might occur, a priori, that a non--trivial map $g\in G$ (for which
$\int \omega_5\neq0$) becomes homotopically trivial if it seen as an
element of $G/H$. Due to eq.~(\ref{eqquantwzw}), the existence of the map
$g$ would imply quantization from the $5D$ point of view, while no
restriction would appear in the standard $4D$ framework. As discussed in
\cite{D'Hoker:1994ti}, however, this can not happen if, as in our case
(\ref{anfh}), the generators which define the CS form provide an
anomaly--free embedding of $H$. In this case, $G$ maps for which $\int
\omega_5\neq0$ are topologically non--trivial in $G/H$ also and then 
 the $5D$ quantization condition coincides with the standard
$4D$ one. In the next section we will provide an explicit example of this
fact, we will apply the $5D$ condition (\ref{eqquantwzw}) to the case of
QCD and show that the result coincides with the well--known $4D$ one.

\subsection{Anomaly in $AdS/QCD$}

It is useful to apply the results of the previous section to the concrete
case of $AdS/QCD$. We consider, as in sect.~4, a gauge group $G = \Gamma_L
\times \Gamma_R$ broken to its diagonal subgroup $H=\Gamma_V$. To apply
the formalism previously outlined, we must choose a representation of $G$
such that the $H$ embedding is anomaly--free (\ref{anfh}). The generators are
\be
T_{a}^L =
\left(
\begin{array}{cc}
t_a & 0\\
0 & 0
\end{array}
\right)\,,
\qquad\qquad
T_{a}^R =
\left(
\begin{array}{cc}
0 & 0\\
0 & - (t_a)^T
\end{array}
\right)
\,,\label{eqGenG}
\ee
where $t_{a}$ denote the generators of $\Gamma$ normalized to
$2{\textrm{Tr}}[t_a,t_b]=\delta_{ab}$.
The vector combinations
\be
T_a =
\left(
\begin{array}{cc}
t_a & 0\\
0 & - (t_a)^T
\end{array}
\right)\,,
\ee
are the generators of the unbroken subgroup $\Gamma_V$.
The gauge field is $A=A_{L}^aT_{a}^L+A_{R}^aT_{a}^R$ and to rewrite our
results with the common notation we also define $A_{L,R} = A^a_{L,R} t^a$.

The CS form is
\be
\omega_5(A) = \omega_5(A_L) - \omega_5(A_R)\,,
\label{oqcd}
\ee
where $\omega_5$ is given in eq.~(\ref{CS}), it is manifest that
$\omega_5(A)$ vanishes for $A \in Lie(\Gamma_V)$, hence $\Gamma_V$ is
anomaly free. The 4-form $B_4$ (eq.~(\ref{eqB4})) which enters in the definition of the
shifted CS (eq.~(\ref{cstl})) is now
\be
B_4 = \frac{1}{2} {\rm Tr}\left[({\cal V} A - A {\cal V})(F + F_V)
+ A {\cal V}^3 - {\cal V} A^3 + \frac{1}{2} {\cal V} A {\cal V} A\right]\,,
\ee
where ${\cal V}=1/2(A_{L}^a+A_{R}^a)T^a$ denotes the vectorial part of the
gauge field, $F \equiv d A + A^2$
is the field strength of the gauge field and $F_V \equiv d{\cal V} + {\cal
V}^2$.
The above equation can be rewritten as
\be
B_4 = \frac{1}{2} {\rm Tr}\left[
({F}_R {A}_R + {A}_R {F}_R) {A}_L
+ {A}_L {A}_R^3 + \frac{1}{4} ({A}_R {A}_L)^2
- (L \leftrightarrow R)\right]\,,
\ee
and coincides with the Adler--Bardeen counterterm \cite{Bardeen:1969md}.

The general gauge--fixing procedure of the previous section
could be directly applied to the present case. This would lead, however,
to a non--standard parametrization of the Goldstone degrees of freedom.
Analogously to what we did in sect.~4, we need to adapt the previous
discussion to the case of $AdS/QCD$. As discussed in appendix~A, the final 
result is the same as eq.~(\ref{gff}) in which $\Lambda$ is now given by
\be
\Lambda (x, z) =
\left(
\begin{array}{cc}
{\widetilde U}^{-1} & 0\\
0 & \I
\end{array}
\right)\,,
\label{eqgamma}
\ee
where $\widetilde U$ is an extension on the disk of the Goldstone boson
matrix $U(x)$, {\it{i.e.}} ${\widetilde U(x,z_{\UV})}=U(x)$, ${\widetilde U(x,z_{\IR})}=\I$.

With $\Lambda$ given in eq.~(\ref{eqgamma}), all the other formulas of the
previous section can be directly applied. The $\chi{\bf PT}$ action
(compare with eq.~(\ref{zgfchiral})) is
\bea
&&\hspace{-5em}\exp\left[i \widetilde S_{\chi{\bf PT}}[l_\mu, r_\mu,
U]\right]
= \exp[i S_{wzw}[l_\mu, r_\mu, U]]\nn\\
&&\hspace{-1em}\times \int {\mc D} A_{L,\mu}(x, z)_{\widehat{A}_{L,\mu} =
l_\mu^{(U)}}
{\mc D} A_{R,\mu}(x, z)_{\widehat{A}_{R,\mu} = r_\mu}
\exp[i \widetilde S[A_\mu, A_z=0]]\,,\label{eqeffactanom}
\eea
where $\widetilde S \equiv S_g + \widetilde S_{CS}$. 
The explicit form of the WZW term can be read from eq.~(\ref{WZWexpl}), it
can be checked that it coincides with the $4D$ result
reported in \cite{Wess:1971yu,Kaymakcalan:1983qq}.
When the sources are set to zero,
eq.~(\ref{WZWexpl}) simplifies and we are left with the usual ungauged
WZW term
\be
S_{wzw}[0, 0, U] = i\,c \int_5 \omega_5[(\widetilde U d \widetilde U^{-1})^5]
=\frac{i\,c}{10} \int_5 {\rm Tr}\left[(\widetilde U d \widetilde
U^{-1})^5\right]\,.
\label{eqWZWungauged}
\ee

As discussed before, when $\pi_5(G)$ is non--trivial the coefficient of
the Chern--Simons term is quantized. In particular, in the
$G = SU(N)_L \times SU(N)_R$ case with $N \geq 3$,
$\pi_5(G) = \ZZ^2$.
The quantization condition can be easily obtained from
eq.~(\ref{eqquantwzw}), $g$ represents in this case the more general map
from $S^5$ to $SU(N)_L\times SU(N)_R$ and $\omega_5$ is given by
eq.~(\ref{oqcd}), notice that its ``$L-R$'' form is due to the
requirement of having an anomaly--free $SU(N)_V$. It is well known that,
for a generic $SU(N)$ element such as $g_L$ or $g_R$, one has
\be
\int_{S^5}\omega_5(g^{-1}dg) = \int_{S^5} \frac{1}{10} {\rm
Tr}\left[\left(g^{-1}\, d g\right)^5\right]
= 48 \pi^3 n\,i\,,
\ee
with $n$ integer. The quantization condition then becomes
$c = m/(24 \pi^2)$ and substituting in eq.~(\ref{eqWZWungauged}) one obtain the usual
quantized form of the WZW term
\be
S_{wzw}[0, 0, U] = \frac{i\,m}{240 \pi^2} \int_5 {\rm Tr}\left[(\widetilde
U d \widetilde U^{-1})^5\right]\,.
\ee
Notice that we obtained the standard quantization condition from purely $5D$ consistency requirements.

\section{Conclusions}

The holographic technique provides a useful tool to deal with $5D$ theories on the interval.
In this paper we have worked out a precise recipe to implement the holographic procedure for a
gauge theory. In particular, we have shown how to include into the holographic action the
degrees of freedom which arise from the $4D$--scalar component of the $5D$ fields. Our main
result is summarized by eq.~(\ref{zgf}), the scalars are described by a $4D$ Goldstone boson
matrix $\Sigma$ and the holographic action is obtained by integrating the $5D$ bulk gauge
fields with $UV$ boundary condition ${\widehat A}=B^{(\Sigma^{-1})}$. We are in the axial
gauge $A_z=0$, so that the $5D$ action has a simplified form and the $IR$ boundary conditions
for $A_\mu(x,z)$ are simply Neumann or Dirichlet for, respectively, the unbroken and broken
generators of the gauge group. It is important to remark that the $G$ invariance of the
holographic action is explicit in the present formalism. The bulk functional integral
is indeed $H$ invariant and the boundary value $\widehat A$ only transforms with $H$ under generic $G$ transformations. In sections 3 and 4 we have considered applications of the holographic method to the Composite--Higgs scenario, showing how to compute the one--loop Higgs potential, and to $AdS/QCD$. We believe we have shown that, in certain cases, the holographic calculation is much more simple than the standard KK one.

At a more theoretical level, we have discussed in section~5 the effect of including a Chern--Simons term in the $5D$ gauge action. We have explicitly shown that it makes an anomaly arise in the $4D$ effective theory and that, as a direct consequence of this fact, a gauged WZW term appears in the holographic Goldstone bosons action. We also pointed out that the ``naive'' CS is not enough to reproduce the anomaly in its standard (shifted, in the language of \cite{Chu:1996fr}) form, one has to add a suitable localized operator and construct a shifted CS term. In $AdS/QCD$, the shifted CS is the only correct choice if one wants to reproduce the Adler--Bardeen chiral anomaly.\footnote{The problem of including anomalies in $5D$ models of $QCD$ has already been discussed in \cite{Hill:2006wu} (see also \cite{Liao:2006qq}). The set--up considered in those papers, however, is quite different from ours. It consists in a non--chiral $\Gamma=SU(3)$ bulk gauge group and two holographic boundaries at which the left-- and right--handed sources are located.} In the framework of $4D$ effective field theories, the WZW term has a quantized coefficient if $\pi_5(G/H)$ is non--trivial, it is worth asking whether the coefficient of the CS term is also quantized for an independent five--dimensional reason. We find that this is
indeed the case, even though the $5D$ quantization condition seems stronger than the $4D$
one as it depends on $\pi_5(G)$ and not on $\pi_5(G/H)$. In $5D$, however, the $H$
embedding in $G$ needs to be anomaly free and this makes the two quantization conditions
coincide, as we explicitly saw to happen in the notable case of $QCD$.

\subsection*{Acknowledgments}
We are grateful to M.~Serone for his help, A.~W. is indebted with A.~Pomarol
for the many useful discussions, G.~P. thanks L.~Bonora for valuable discussions.
We also thank R.~Contino, M.~Quir\'os, A.~Pomarol and M.~Serone for giving us their
opinion on the manuscript.

\appendix

\section{The Holographic Gauge--Fixing}

Let us start from the holographic partition function (\ref{partf}), we want to gauge--fix its local ${\mc G}_B$ invariance. Looking at eq.~(\ref{one}) we see that it can not be used to fix
the gauge as the $g$ elements one integrates over do not reduce to $H$ at
the $IR$. It is then convenient to rewrite the integral on the group
isolating the integral on the $IR$ variables, and again split the latter
in integrals over $H$ and $G/H$.\footnote{If $G$ is a compact Lie group
and $H$ a closed subgroup, the group integral can be split as \cite{Weil}
$$
\int_Gdg\,f[g]=\int_{G/H}d\gamma\,\{\int_Hdh\,f[\gamma\circ h]\}\,.
$$
where the appropriate left invariant Haar measures are understood in all integrals.}
Be $F$ any functional on the local group, we have
\bea
&&\int{\mc D}g(x,z)_{\widehat{g}=\I}F[g]=\int{\mc
D}\widetilde{g}(x)\int{\mc
D}g(x,z)_{\widehat{g}=\I,\,g_{IR}=\widetilde{g}}F[g]\nn\\
&&=\int_{G/H}{\mc D}\widetilde{\gamma}(x)\int_H{\mc
D}\widetilde{h}(x)\int{\mc
D}g(x,z)_{\widehat{g}=\I,\,g_{IR}=\widetilde{\gamma}\circ\widetilde{h}}F[g]\,,\label{eqgf1}
\eea
where $\widetilde{\gamma}$ is an element the right $G/H$ coset,
$\widetilde{h}\in H$ and the left Haar measure is understood in all group
integrals. We can now remove the $G/H$ component of $g_{IR}$ by performing
a change of variable in the $g$ integral with a $5D$ transformation
$\widetilde{\Lambda}$ such that $\widetilde{\Lambda}(z_{\UV})=\I$ and
$\widetilde{\Lambda}(z_{\IR})=\widetilde{\gamma}$.\footnote{We are assuming
that $\pi_4(G/H)$ is trivial (see Footnote~4) so that the existence of
$\widetilde{\Lambda}$ is guaranteed.} After the change of variable,
$g_{IR}=\widetilde{h}$ belongs to $H$ so that when putting together the
${\mc D}\widetilde{h}(x)$ and ${\mc D}g(x,z)$ integrals we obtain an
integral over the bulk gauge group ${\mc G}_B$, {\it{i.e.}} with the
correct restriction at the $IR$. Then, we have
\be
\int{\mc D}g(x,z)_{\widehat{g}=\I}F[g]=\int_{G/H}{\mc
D}\widetilde{\gamma}(x)\int_{{\mc G}_B}{\mc
D}g(x,z)_{\widehat{g}=\I}F[\widetilde{\Lambda}\circ g]\,.
\label{gint}
\ee
Note that $\widetilde{\gamma}(x)$ can be identified with the Goldstone
boson $\Sigma$ we defined in sect.~\ref{holgaugefix}, so that we will
change name to it from now on.\footnote{To ensure $G$--invariance, the
measure ${\mc D}\Sigma$ for the Goldstone bosons must be the left invariant Haar
measure of $G/H$, and this is exactly what we will find. The explicit form of
${\mc D}\Sigma$ is in general non--trivial, and could be relevant for
particular applications. A brief discussion of this subject and references
to the original literature can be found in \cite{Leutwyler:1993iq}.}

Let us now multiply the partition function (\ref{partf}) by ``$1$'' written as
in eq.~(\ref{one}) and use eq.~(\ref{gint}) to rewrite the group integral.
Changing variables in the gauge field integral we get
\be
Z\left[B_\mu\right]\,=\, \int_{{\mc G}_B}{\mc D}g\int{\mc D}\Sigma\int
\mathcal{D}A_\mu(x,z)_{\widehat{A}_\mu=B_\mu}\exp{\left[iS\left[\left(A^{(\widetilde{\Lambda}^{-1})}\right)^{(g^{-1})}\right]\right]}\,,
\label{gf2}
\ee
where we used the delta function to perform the $A_z$ integral, so that in
the above equation $A$ indicates a $5D$ vector with zero fifth component.
Moreover, we neglected the determinant of eq.~(\ref{one}) and did not
specify the $IR$ boundary condition of the $\mathcal{D}A_\mu$ integral.
After the $\widetilde{\Lambda}$ transformation those are given by
eq.~(\ref{bcn}). Our action, including the CS term, is by
definition invariant under the elements of ${\mc G}_B$ since they reduce
to $H$ at the $IR$. The ${\mc G}_B$ integral then factorizes out in
eq.~(\ref{gf2}), we drop it and finally obtain the gauge--fixed partition
function.  Ordinary $IR$ boundary conditions (\ref{bc}) are restored by
changing variable in the $\mathcal{D}A_\mu$ integral with a $5D$ gauge
transformation of parameter $\Sigma(x)$. The result is reported in eq.~(\ref{gff}).

In the case of $AdS/QCD$ a slight change in the above derivation is needed in order to match the standard parametrization of the Goldstone boson fields. The analog of eq.~(\ref{eqgf1}) reads
now
\bea
&&\int {\mc D} g_\Rs(x,z) {\mc D} g_\Ls(x,z) F[g_\Ls, g_\Rs]\nn\\
&&= \int {\mc D} \widetilde g_\Rs(x) {\mc D} \widetilde g_\Ls(x)
\int {\mc D} g_\Rs(x,z)_{g_\Rs(z_\IR) = \widetilde g_\Rs}
{\mc D} g_\Ls(x,z)_{g_\Ls(z_{\IR}) = \widetilde g_\Ls} F[g_\Ls, g_\Rs]\nn\\
&&= \int {\mc D} \widetilde g_\Rs(x) {\mc D} \widetilde \gamma(x)
\int {\mc D} g_\Rs(x,z)_{g_\Rs(z_\IR) = \widetilde g_\Rs}
{\mc D} g_\Ls(x,z)_{g_\Ls(z_{\IR}) = \widetilde \gamma \circ \widetilde
g_\Rs} F[g_\Ls, g_\Rs]\,,
\eea
where we performed a change of variable
$\widetilde g_\Ls(x) \rightarrow \widetilde \gamma \circ \widetilde
g_\Rs(x)$ with $\widetilde \gamma \in \Gamma$ and omitted for shortness the $UV$ restriction ${\widehat g}_{L,R}=\I$.
Changing variables in the $g_L$ integral by a $5D$ transformation
$\widetilde \Lambda$ such that $\widetilde \Lambda(z_\IR) = \widetilde \gamma$
and  $\widetilde \Lambda(z_\UV) = \I$ we get
\bea
&&\int {\mc D} g_\Rs(x,z)_{\widehat g_\Rs = \I} {\mc D}
g_\Ls(x,z)_{\widehat g_\Ls = \I} F[g_\Ls, g_\Rs]\nn\\
&&= \int {\mc D} \widetilde \gamma(x) \int {\mc D} \widetilde g_\Rs(x)
\int {\mc D} g_\Rs(x,z)_{g_\Rs(z_\IR) = \widetilde g_\Rs}
{\mc D} g_\Ls(x,z)_{g_\Ls(z_{\IR}) = \widetilde g_\Rs} F[\widetilde \Lambda
\circ g_\Ls, g_\Rs]\,.
\eea
Putting the $\widetilde g_\Rs$ and $g_{\Rs, \Ls}$ integrals together we
obtain an integral
over the ${\mc G}_B$ bulk gauge group.
The matrix $\widetilde \gamma \in \Gamma$ can be identified with the
Goldstone boson
matrix $U^{-1}$ of $\chi{\bf PT}$.
We can now follow the procedure used in the general case and we get the same as eq.~(\ref{gff}) in which $\Lambda$ is now given by eq.~(\ref{eqgamma}).

\section{Holography with Boundary Terms}\label{appBoundaryTerms}

In this appendix we will briefly discuss how to derive the holographic effective action in the presence of localized terms at both the $IR$ and $UV$ branes. Such terms appear in many interesting situations. In the presence of bulk scalars
in $AdS_5$, for instance, localized masses are needed in order to have a zero mode. Furthermore, boundary terms for fermions or gauge fields as well as fields localized on the $UV$ boundary are often present in realistic models both in flat \cite{Scrucca:2003ra, Panico:2006em} and in warped space \cite{Agashe:2004rs,Agashe:2005dk,Contino:2006qr}.

The inclusion of localized $IR$ terms in the holographic procedure is simply accomplished by modifying the $IR$ boundary conditions.\footnote{Notice that the inclusion of $IR$ boundary terms may induce a breaking of $G/H$ at that boundary
directly into the action and not only through the boundary conditions. In such
case the correct gauge--fixing procedure is the one described in appendix~A.} The latter are indeed modified by the boundary terms and are derived requiring the variation of the action on the
IR boundary to vanish. Once the bulk equations of motion are solved with
these new $IR$ conditions, the tree--level effective action can be computed as
in section~3 by plugging the solutions back into the $5D$ action.

The treatment of $UV$ localized terms is even more straightforward.
These terms are functions of the field values at the $UV$ boundary,
thus they are functions of the holographic fields
and can be directly inserted into the effective action. This is clearly true for
gauge and scalar fields, indeed in such cases one always chooses
the boundary values of the non--vanishing fields at the $UV$ boundary
as holographic degrees of freedom
and a localized term can be simply translated into a term of the effective action
\be
S_{loc}^{UV} = \int d^4x \int dz\ \delta(z-z_{\UV}) {\cal L}(A_\mu, \Phi)
\rightarrow S_{h} = \int d^4x\ {\cal L}(B_\mu, \phi)\,,
\ee
where $B_\mu$ and $\phi$ are the sources corresponding respectively
to the gauge field $A_\mu$ and to the scalars $\Phi$.

A slightly different approach must be used when localized $UV$ terms for the fermions
are present. In this case left- and right-handed components are related by the
bulk equations of motion and only one of the two can be chosen as holographic degree
of freedom. Usually one chooses holographic fields of only one chirality, say L,
independently of the actual $UV$ boundary conditions of the fields. In this case, as
explained in section~\ref{secEffAct}, one introduces some Lagrange multipliers in the
effective action in order to set to zero the components with D $UV$ boundary conditions.

In the presence of localized terms for a component which is non--vanishing at
the $UV$ boundary two possibilities can arise.
As a first case, if that component has been chosen as a holographic degree of freedom, the
localized terms can be directly included in the effective action as we did for the
bosonic fields
\be
S_{loc}^{UV} = \int d^4x \int dz\ \delta(z-z_{\UV}) {\cal L}(\psi_L)
\rightarrow S_{h} = \int d^4x\ {\cal L}(\chi_L)\,,
\ee
where $\chi_L$ is the source related to $\psi_L$.
Otherwise, if the boundary term is a function of a component which is not present among the
holographic fields, one can include it into the effective action using the corresponding
Lagrange multiplier (see \cite{Contino:2004vy} for a more
detailed discussion)
\be
S_{loc}^{UV} = \int d^4x \int dz\ \delta(z-z_{\UV}) {\cal L}(\psi_R)
\rightarrow S_{h} = \int d^4x\ {\cal L}(\lambda_R)\,,
\ee
where $\lambda_R$ is the Lagrange multiplier associated to $\psi_R$.

Another situation which can be easily handled within the holographic procedure is
when localized fields at the $UV$ boundary are coupled with bulk fields. The action for the localized fields can be directly introduced into
the effective action
once their couplings with the bulk fields are rewritten in terms of the
sources (or, if necessary, in terms of the Lagrange multipliers).

\subsection{Effective action for Scalars with Localized Mass Terms}

To clarify the treatment of boundary terms within the holographic procedure, let us derive the effective action for a bulk scalar field $\Phi$, in a certain representation of the bulk group $G$, with localized mass terms. The bulk action is given by
\be
S_5 = \frac{1}{g_5^2} \int d^4x \int_{z_{\UV}}^{z_{\IR}} dz\ \sqrt{g}\,
\left[\frac{1}{2}D_M \Phi^\dagger D^M \Phi
-\frac{1}{2} m_\phi^2 \Phi^\dagger \Phi\right]\,.
\label{eqActionScalars}
\ee
For the components which are non--vanishing at the boundaries
we also add localized mass terms
\be
S_{loc} = \frac{1}{2 g_5^2} \int d^4x \int_{z_{\UV}}^{z_{\IR}}\!\!\! dz\, a^4(z)
\left\{\delta(z-z_{\IR})\sum_a b_{\IR}^a \Phi^{a \dagger} \Phi^a - \delta(z-z_{\UV}) \sum_{a'} b_{\UV}^{a'} \Phi^{{a'} \dagger} \Phi^{a'}\right\}\,,
\label{eqloctermsapp}
\ee
where the sum indices $a$ ($a'$) run over the components which are non--vanishing at the
IR (UV) boundary.

The holographic procedure, in analogy with the case of gauge and fermion fields (compare
eq.~(\ref{zgffer})),
determines a relation between the $UV$ value of the bulk fields and the sources
\be
\widehat\Phi^A(x) = (\phi^{(\Sigma_s^{-1})})^A(x)\,,\label{eqscalarsource}
\ee
where $\phi$ represents the dynamical sources on the $UV$ boundary,
and $\Sigma_s$ is the Goldstone boson matrix in the appropriate $G$ representation.
To compute the quadratic effective action we must solve the
linearized bulk EOM's (see eq.~(\ref{eqscalEOM})) and express the solutions
as functions of the values of the fields on the $UV$ boundary
\be
\Phi^A(p,z) = \widehat\Phi^A(p) f_s^A(p^2, z)\,,\label{eqscalarsolut}
\ee
with $f_s^A(p^2, z_{\UV}) = 1$. The $IR$ boundary conditions induced by the
boundary terms are
\be
\left\{
\begin{array}{l}
\left.\left[\partial_z \Phi^a(p, z) - a(z) b^a_{\IR} \Phi^a(p, z)\right]\right|_{z=z_{\IR}} = 0\,,\\
\Phi^{\widehat a}(p, z_{\IR}) = 0\,.
\end{array}
\right.
\ee
Substituting the solutions into the action and using eq.~(\ref{eqscalarsource}) we find
the tree level holographic action
\be\label{eqHolActScalar1}
S_{h} = \frac{1}{2 g_5^2}\int d^4 x \sum_A \left(\phi^{\dagger}\, \Sigma_s\right)^A\, \Pi_s^A(-\partial^2)
\left(\Sigma_s^\dagger \phi\right)^A- \frac{1}{2 g_5^2} \int d^4x \sum_a b_{\UV}^a \phi^{a \dagger} \phi^a\,,
\ee
where $\Pi_s^A(p^2) = \partial_z f_s^A(p^2,z)|_{z={z_{\UV}}}$.

The computation of the contribution to the one--loop effective potential coming
from scalar fields
is similar to the one described for gauge bosons and fermions, thus we will briefly
report the results.
The quadratic effective action for a scalar field, in the presence of a
background $\ov \Sigma$ for the Goldstone bosons, can be written as
\be
S_{h} = \frac{1}{g_5^2}\int d^4 x\ \sum_{a,b} \phi^{a\,\dagger} \Pi_s^{a,b}(\ov\Sigma)\,\phi^b\,,
\ee
where the sum is over the dynamical fields.
The scalar contribution to the one--loop effective potential is
\be
V_s(\ov \Sigma) = \int \frac{d^4p_E}{(2\pi)^4}
\log\left[{\rm Det}\left(\Pi_s(\ov\Sigma)\right)\right]\,,
\ee
where the momentum has been rotated into the Euclidean space.

\section{Bulk Wavefunctions}\label{appSolEOM}

In this Appendix we report the form of the solutions of the bulk EOM's, which are
needed to compute the holographic action and the effective potential. We derive
the wavefunctions for gauge fields, fermions and scalars in the AdS and flat space cases.
For gauge fields and fermions we consider an action without localized terms.
In the case of scalars, as customary, we include also the boundary
mass terms which are needed to obtain $4D$ zero modes in the AdS case.

\subsection{Gauge Fields}\label{appGauge}

We consider a gauge theory with gauge group $G$ broken to a subgroup $H$ at
the $IR$ boundary. We will denote by $a$ the unbroken generators, namely
those in $H$, and by $\widehat a$ the broken generators, namely those in $G/H$.
The $5D$ action is given by
\be
S_B = \frac{1}{g_5^2}\int d^4 x \int_{z_{\UV}}^{z_{\IR}} dz\ a(z)\ {\rm Tr}\left[
-\frac{1}{2} F_{\mu\nu} F^{\mu\nu} + \partial_z A_\mu \partial_z A^\mu\right]\,.
\ee
The bulk EOM's are
\be
\partial_\mu \partial^\mu A_{t,\nu}^a
- \frac{1}{a(z)} \partial_z (a(z) \partial_z) A_{t,\nu}^a = 0\,,\;\;\;\;\;-\frac{1}{a(z)} \partial_z (a(z) \partial_z) A_{l,\nu}^a = 0\,,
\label{eqeqofmotgauge}
\ee
where $A_{t,l}^\mu$ are, respectively, the transverse and longitudinal components. The $IR$ boundary conditions are
\be
\left\{
\begin{array}{l}
\left.\partial_z A_{t,l\,\nu}^{a}\right|_{z=z_{\IR}} = 0\,,\\
\left.A_{t,l\,\nu}^{\widehat a}\right|_{z=z_{\IR}} = 0\,.
\end{array}
\right.
\ee
In the following we will denote generically by $A^+$ and $A^-$ any field corresponding
respectively to the unbroken subgroup $H$ or to the broken generators in $G/H$.

Using the definitions in eq.~(\ref{eqFieldExpansion}) and in eq.~(\ref{eqscalwfsimpl})
in the AdS case ($a(z) = L/z$) we find
\be\label{eqGaugeEOM}
p^2 F^\pm(p, z) + z \partial_z \left(\frac{1}{z} \partial_z \right) F^\pm(p, z) = 0\,,
\ee
with the $IR$ boundary conditions
\be
\left\{
\begin{array}{l}
\left.\partial_z F^+(p, z)\right|_{z=z_{\IR}} = 0\,,\\
\left.F^-(p, z)\right|_{z=z_{\IR}} = 0\,.
\end{array}
\right.
\ee
At the $UV$ boundary we must impose
\be
F^\pm(p, z_{\UV}) = 1\,.
\ee
The general solution of eq.~(\ref{eqGaugeEOM}) is
\be
F^\pm(p, z) = b\, z (J_1(p z) + c\, Y_1(p z))\,.
\ee
Imposing the boundary conditions we get
\be
F^\pm(p, z) =\displaystyle \frac{h_\pm(p, z)}{h_\pm(p, z_{\UV})}\,,
\label{eqSolGauge1}
\ee
with
\be
\left\{
\begin{array}{l}
h_+(p, z) = z \left(Y_0(p z_{\IR}) J_1(p z) - J_0(p z_{\IR}) Y_1(p z)\right)\,,\\
h_-(p, z) = z \left(Y_1(p z_{\IR}) J_1(p z) - J_1(p z_{\IR}) Y_1(p z)\right)\,.
\end{array}
\right.\label{eqSolEOMGaugeAdS}
\ee
The derivatives of $h_\pm$ with respect to $z$ are
\be
\left\{
\begin{array}{l}
h'_+(p, z) = p z \left(Y_0(p z_{\IR}) J_0(p z) - J_0(p z_{\IR}) Y_0(p z)\right)\,,\\
h'_-(p, z) = p z \left(Y_1(p z_{\IR}) J_0(p z) - J_1(p z_{\IR}) Y_0(p z)\right)\,,
\end{array}
\right.
\ee
so that
\be
\Pi_t^\pm =\displaystyle \frac{h'_\pm(p, z_{\UV})}{h_\pm(p, z_{\UV})}\,,
\label{eqSolGauge2}
\ee
A simple relation links $\Pi_t^A$ to $\Pi_t^0$ and $\Pi_t^1$:
\be
\Pi_t^\pm =\displaystyle \Pi_t^0 \pm \Pi_t^1\,.
\label{eqSolGauge3}
\ee

For a theory compactified on a flat space ($a(z)=1$, $z_{\UV}=0$ and $z_{\IR} = \pi R$),
the definitions in eqs.~(\ref{eqSolGauge1}), (\ref{eqSolGauge2}) and (\ref{eqSolGauge3})
remain valid and $h_\pm$ are given by
\be
\left\{
\begin{array}{l}
h_+(p, z) = \cos(p(\pi R - z))\,,\\
h_-(p, z) = \sin(p(\pi R - z))\,.
\end{array}
\right.
\ee

\subsection{Fermions}\label{appFermions}

We consider a bulk fermion $\Psi$ with an odd mass term
\be
S_5 =  \frac{1}{g_5^2} \int d^4x \int_{z_{\UV}}^{z_{\IR}} dz\ \sqrt{g}\, \left[
\frac{i}{2} \ov \Psi e^M_A \Gamma^A D_M \Psi
- \frac{i}{2} (D_M \Psi)^\dagger \Gamma^0 e^M_A \Gamma^A \Psi
-M \ov \Psi \Psi\right]\,,
\label{eqFermionsAct}
\ee
where we defined the $5D$ $\Gamma$ matrices $\Gamma^A = \{\gamma^\mu, -i \gamma^5\}$,
the inverse vielbein $e^M_A=\delta^M_A/a(z)$ and the covariant derivative
$D_M = \partial_M + \frac{1}{8} \omega_{\scriptscriptstyle{MAB}} [\Gamma^A, \Gamma^B]$ with
$\omega_{\scriptscriptstyle{MAB}}$ whose only non vanishing component is
$\omega_{\mu a5} = (\eta_{\mu a}/a(z))\partial_z a(z)$.
Notice that we normalized the action with a factor
$1/g_5^2$, where $g_5$ is the $5D$ gauge coupling.
This ensures that the dimensions of the fields are the same in $5D$ and in the
$4D$ effective action and simplifies the relation between bulk and holographic fields.
The bulk equations of motion are
\be
\left[ \partial_z + 2 \frac{\partial_z a(z)}{a(z)} \pm a(z) M\right] \psi_{L,R}
= \pm \pslash\, \psi_{R,L}\,,
\label{eqfermEOM}
\ee
where $\psi_{R,L}$ denote the left-- and right--handed components of $\Psi$ which satisfy
the condition $\gamma^5 \psi_{R,L} = \pm \psi_{R,L}$.
The bulk equation can be solved rewriting $\psi_{R,L}$ as
\be
\psi_{R,L} (p,z) = d_{R,L}(p,z) \widehat\psi_{R,L}(p)
\equiv \frac{h_{R,L}(p,z)}{h_{R,L}(p,z_{\UV})} \widehat\psi_{R,L}(p)\,,
\ee
where $\widehat\psi_{R,L}$ are the values of the fields on the $UV$ boundary and $d_{R,L}$
satisfy the condition $d_{R,L}(p,z_{\UV}) = 1$. The equations of motion relate
$\widehat\psi_{R}$ and $\widehat\psi_{L}$:
\be
\pslash\, \widehat\psi_{R}(p) = p\,\frac{h_{R}(p,z_{\UV})}{h_{L}(p,z_{\UV})} \widehat\psi_{L}(p) = p\, f_R(p, z_\UV) \widehat \psi_L(p)\,,
\ee
so that only one of the two fields can be chosen as holographic degree of freedom.

In the AdS case, the bulk equations of motion become
\be
\left[z\partial_z - (2 \mp ML)\right] h_{L,R}(p,z)
= \pm z\, p\, h_{R,L}(p,z)\,.
\ee
Denoting by $d^\pm$ the solutions
in the cases in which $\psi_{R,L}$ has D boundary conditions at the $IR$ one finds
\be
\left\{
\begin{array}{l}
h_L^\pm(p, z) = z^{5/2} \left[Y_{c\mp1/2}(p z_{\IR}) J_{c+1/2}(p z)
- J_{c\mp1/2}(p z_{\IR}) Y_{c+1/2}(p z)\right]\,,\\
h_R^\pm(p, z) = z^{5/2} \left[Y_{c\mp1/2}(p z_{\IR}) J_{c-1/2}(p z)
- J_{c\mp1/2}(p z_{\IR}) Y_{c-1/2}(p z)\right]\,,
\end{array}
\right.
\ee
where $c \equiv ML$.

In the flat space case one finds
\be
\left\{
\begin{array}{l}
h_L^-(p, z) = \displaystyle\sin(\omega(\pi R - z))\,,\\
h_R^-(p, z) = \displaystyle\frac{1}{p}\left[-\omega \cos(\omega(\pi R - z)) + M \sin(\omega(\pi R - z))\right]\,,
\end{array}
\right.
\ee
and
\be
\left\{
\begin{array}{l}
h_L^+(p, z) = \displaystyle\frac{1}{p}\left[\omega \cos(\omega(\pi R - z)) + M \sin(\omega(\pi R - z))\right]\,,\\
h_R^+(p, z) = \displaystyle\sin(\omega(\pi R - z))\,,
\end{array}
\right.
\ee
where $\omega^2 = p^2 - M^2$.

\subsection{Scalars}\label{appScalars}

We consider a bulk scalar field $\Phi$ with bulk action given in eq.~(\ref{eqActionScalars}).
For a field which is non--vanishing at both boundaries we also add localized mass terms
of the form
\be
S_{loc} = \frac{1}{2 g_5^2} \int d^4x \int_{z_{\UV}}^{z_{\IR}} dz\ a^4(z)\,
\left\{-b \left[\delta(z-z_{\UV}) - \delta(z-z_{\IR})\right]\Phi^\dagger \Phi\right\}\,,
\label{eqloctermsscal}
\ee
where $b=(2\pm\alpha)/L$ and $\alpha = \sqrt{4 + m_\phi^2 L^2}$.
This particular choice of the coefficients is the one which allows a $4D$ scalar zero mode
for an $AdS$ metric.
Of course when the scalar field satisfies D conditions at a boundary the corresponding
mass term in eq.~(\ref{eqloctermsscal}) is not included.

In this appendix we are interested in computing the solutions of the bulk EOM's with
fixed $UV$ values for the fields, thus we only need to consider the $IR$ localized
terms. In order to solve the linearized equations of motion we write the fields as
\be
\Phi(p, z) = \widehat\Phi(p) f(z)\,.
\ee
Imposing the variation of the action in eq.~(\ref{eqActionScalars}) to vanish, we find
\be
p^2 f(z) + \frac{1}{a^3(z)} \partial_z(a^3(z) \partial_z f(z))
+ a^2(z) m_\phi^2 f(z) = 0\,,\label{eqscalEOM}
\ee
while the $IR$ boundary conditions are
\be
\left\{
\begin{array}{l}
(a^{-1}(z) \partial_z - b)f^+(z)|_{z=z_{\IR}} = 0\,,\\
f^-(z)|_{z=z_{\IR}} = 0\,,
\end{array}
\right.
\ee
where $f^\pm$ represent the solutions which are respectively non--vanishing at the $IR$ boundary
or satisfy D conditions.

In the AdS case the solutions of the equations of motions are
\be
f^\pm(p,z) = \frac{h^\pm(p,z)}{h^\pm(p,z_{\UV})}\,,
\label{eqparscal}
\ee
where
\be
\left\{
\begin{array}{l}
h^+(p,z) = z^2\left[Y_{\alpha\pm1}(p z_{\IR}) J_\alpha(p z) - J_{\alpha\pm1}(p z_{\IR}) Y_\alpha(p z)\right]\,,\\
h^-(p,z) = z^2\left[Y_\alpha(p z_{\IR}) J_\alpha(p z) - J_\alpha(p z_{\IR}) Y_\alpha(p z)\right]\,,
\end{array}
\right.
\ee
with the $\pm$ signs corresponding to the cases $b = (2 \pm \alpha)/L$.

In the flat space case with arbitrary $b$, the solutions of the equations of
motion can be written as in eq.~(\ref{eqparscal}) with
\be
\left\{
\begin{array}{l}
h^+(p,z) = \displaystyle\cos(\omega(\pi R - z)) - \frac{b}{\omega}\sin(\omega(\pi R - z))\,,\\
h^-(p,z) = \sin(\omega(\pi R - z))\,,
\end{array}
\right.
\ee
where $\omega^2 = p^2 - m_{\phi}^2$.

\begin{small}
\providecommand{\href}[2]{#2}\begingroup\raggedright\endgroup

\end{small}

\end{document}